\documentclass[aps,amsfonts,prx,twocolumn,showpacs]{revtex4-1}

\usepackage{subfigure}
\usepackage{wrapfig}
\usepackage{textcomp}
\usepackage{graphicx}
\usepackage{amssymb}
\usepackage{amsmath}
\usepackage{bm}
\usepackage{amsmath, amsthm, amssymb}
\usepackage{hyperref}

\usepackage{amsfonts}
\usepackage{dcolumn}
\usepackage{float}
\usepackage{bm}

\def\beq{\begin{equation}}
\def\eeq{\end{equation}}
\def\bea{\begin{eqnarray}}
\def\eea{\end{eqnarray}}

\newcommand{\cH}{{\cal H}}

\newcommand{\lt}{<}

\def\der#1#2{{\partial#1 \over \partial#2}}
\def\be{\begin{equation}}
\def\ee{\end{equation}}
\def\ba#1{\begin{array}{#1}}
\def\ea{\end{array}}
\def\bn{\begin{enumerate}}
\def\en{\end{enumerate}}

\def\rr{\right}
\def\l{\left}

\def\summ{\sum\limits}

\def\H{{\cal H}}
\def\cF{{\cal F}}
\def\intt{\int\limits}
\def\sx{\sigma_x}
\def\sz{\sigma_z}
\def\sy{\sigma_y}

\def\bmize{\begin{itemize}}
\def\emize{\end{itemize}}

\def\ket#1{\left |#1\right\rangle}
\def\bra#1{\left \langle #1\right|}
\def\braket#1{\left \langle #1\right\rangle}

\begin{document}

\title{
Topological frequency conversion in strongly driven quantum systems
}

\author{Ivar Martin$^{1,2}$, Gil Refael$^{3,4}$, and Bertrand Halperin$^5$}

\affiliation{$^1$ Materials Science Division, Argonne National Laboratory, Argonne, Illinois 60439, USA\\
$^2$ Kavli Institute for Theoretical Physics, University of California, Santa Barbara, CA 93106, USA\\
$^3$ Institute of Quantum Information and Matter and Department of Physics\\
$^4$ Walter Burke Institute of Theoretical Physics, California Institute of Technology, Pasadena, CA 91125 USA\\
$^5$ Department of Physics, Harvard University, Cambridge MA 02138, USA}

\date{\today}

\begin{abstract}

When a physical system is subjected to a strong external multi-frequency drive, its dynamics can be conveniently represented in the multi-dimensional Floquet lattice. 
The number of the Floquet lattice dimensions equals  the number of {\em irrationally}-related drive frequencies, and the evolution occurs in response to a built-in effective  ``electric" field, whose components are proportional to the corresponding drive frequencies.    The mapping allows to engineer and study temporal analogs of many real-space phenomena.
Here we focus on the specific example of a two-level system under two-frequency drive that induces topologically nontrivial band structure in the 2D Floquet space.  The observable consequence of such construction is quantized  pumping of energy between the sources with frequencies $\omega_1$ and $\omega_2$. 
When the system is initialized into a Floquet band with the Chern number $C$, the pumping occurs at the rate $P_{12} = -P_{21}= (C/2\pi)\hbar \omega_1\omega_2$, 
an exact counterpart of the transverse current in a conventional topological insulator.

\end{abstract}

\maketitle

\section{Introduction}

A major goal of quantum condensed matter physics is the control of many-body electronic and atomic states. A periodic drive is emerging as one of the most exciting means for achieving such control. Many proposals for phases that could be induced by periodic drive, so-called Floquet phases,  have been made recently. Refs. \cite{ Oka-09, Inoue10,FTI, FTI-3d, Kitagawa} demostrate driving a band insulator into a topological phase using circularly polarized radiation, or an alternating Zeeman field.  Refs. \cite{AFAI-1,AFAI-2,Lindner11, NathanRudner} explored topological invariants unique to periodically driven phases, and predicted the Anderson-Floquet Anomalous Insulator, an unusual system with fully localized bulk but protected edge states. A general invariant for interacting systems was proposed in \cite{Roy1,Roy2,Sondhi,potter}. Refs. \cite{TC1,TC2,TC4,TC5} even showed that a driven disordered system in 1D could spontaneously break the discrete time translation symmetry, forming the long sought-after time crystal, while Refs. \cite{Abanin, Roderich-FL,Huse} showed that a periodic drive can delocalize a many body localized system. This long list is a clear indication for the richness of Floquet engineering, with some of the proposals already having attracted nascent experimental efforts \cite{Gedik, Rechtsman2013, Monroe}.

A periodic drive alters the form of a quantum wave function. Each state becomes dressed by all possible harmonics of the drive frequency. The extra degrees of freedom associated with the amplitudes of the various drive harmonics effectively raise the dimensionality  of the system, and allow it to exhibit new phenomena. While this observation is at the basis of some of the work mentioned above, the role of this extra emergent dimension remains little utilized, and, even worse, little understood. 

Interestingly, there is another example where  an extra dimensions ``magically" emerge: quasicrystals. Quasicrystals are aperiodic structures that could be understood as projections of higher dimensional periodic crystals onto lower dimensions \cite{KitaevLevitov, LevineSteinhardt, Shechtman}. Most simply, a 1-dimensional (1D) quasicrystal can be constructed by superimposing two periodic but mutually incommensurate potentials.

\begin{figure}
a.\includegraphics[width=0.95\columnwidth]{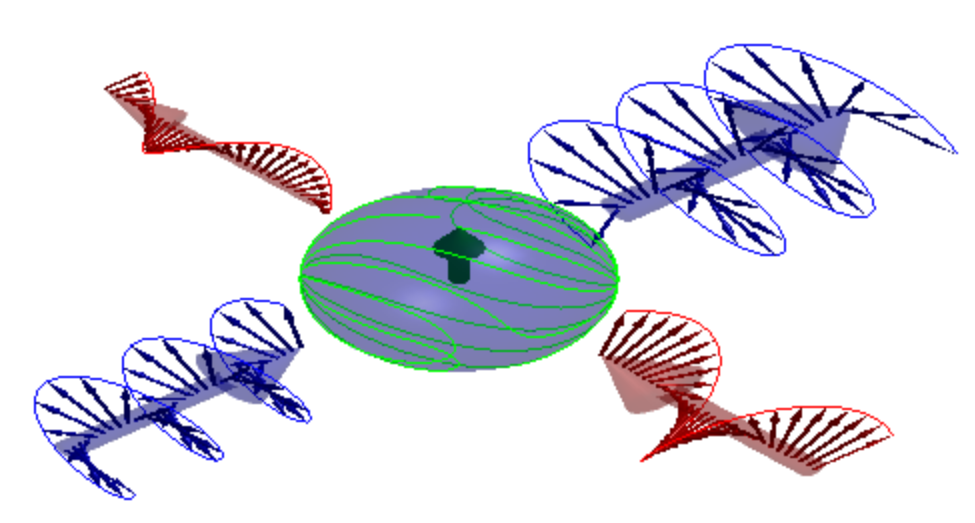}
b.\includegraphics[height=0.2\columnwidth]{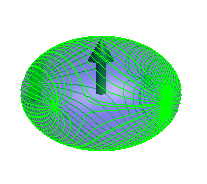}
c.\includegraphics[height=0.2\columnwidth]{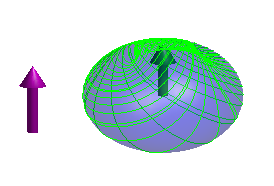}
d.\includegraphics[height=0.2\columnwidth]{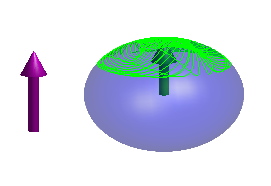}
\caption{In this manuscript we present a model that uses 2D topological insulator band structure as a guide to engineering the quantized energy pumping between different frequency modes.  (a) A spin 1/2 driven by two sources with incommensurate frequencies. In the topological phase, the spin trajectory (green line) fully covers the Bloch sphere (blue). The intermode coupling induced by the spin dynamics pumps energy from one drive (red) to the other (blue) at a nearly quantized rate.  Lower panel shows effective magnetic field trajectories for (b) $m = 0$ (gapless phase between two topological),(c) $m = 1.8$ (just inside topological phase), and  (d) $m = 2.2$ (just outside topological phase). The mass $m$ is depicted as a purple arrow (static $B_z$ field, see Eq. (\ref{2f}) for the definition). 
\label{front-fig}}
\end{figure}

Here we show a surprising consequence of  combining these two schemes for increasing the dimensionality of a system. 
The number of extra {\em} time dimensions is given by the number of the applied drives with incommensurate frequencies. This can give rise to topological phenomena rooted solely in the time dimension.  In particular, we demonstrate that  subjecting a single spin-1/2 particle to two elliptically polarized periodic waves  can realize the chiral Bernevig-Hughes-Zhang (BHZ) model \cite{Bernevig06}, which usually resides in two spatial dimensions, see Fig. \ref{front-fig}.  

What is the consequence of such a construction? Just as edge states are the earmarks of spatial topological phenomena, the signature of temporal topological phenomena arising from incommensurate drives is pumping. We will show that by combining drives into a topological temporal texture, the system will pump energy between the driving fields, drawing energy from one, and feeding it into the other. This general principle could be used, for instance, to convert photons between incommensurate photonic modes in optical cavities (in contrast to the perturbative nonlinearities that lead to the more conventional  phenomena, such as frequency halving or doubling, e.g. \cite{Muka}). In addition to establishing the pumping properties of the topological drive textures, we will demonstrate the existence of Floquet eigenstates for quasiperiodic drives, despite the lack of temporal periodicity. These states can be obtained through a convergent limiting procedure of approximating irrational frequency ratio by rational numbers.

Our paper is organized as follows. After providing some background on Floquet theory, and the physics of the Wannier-Stark Ladder, we move on to consider the general properties of a doubly-driven system in Sec. \ref{sec:2w}. Next, in Sec. \ref{sec:BHZ}, we introduce the BHZ path to temporal topological systems, and explore the model's properties.  In Sec. \ref{nums} we explore the temporal BHZ model numerically, and show that the topological regime is characterized by energy pumping. In Sec. \ref{disc} we discuss the connection of our results with other systems and recent discoveries, and in Sec. \ref{sec:out} we briefly discuss some other directions that can be pursued by means of multi-frequency Floquet engineering.

\section{Background}

\subsection{Floquet theory and the Floquet lattice}

The case of a simple periodic drive of arbitrary strength can be conveniently treated by the Floquet theory \cite{Floquet}, which is a time-analog of the Bloch theory for particles in spatially periodic potentials. In Bloch theory \cite{Bloch}, particles carry quasi-momentum; analogously, in Floquet theory, energy of eigenstates is replaced by quasi-energy $E$, and the Floquet quasi-eigenstates  have the form $\Psi(t) = e^{-iEt} \Phi(t)$. When a system is driven by single frequency $\omega$ (and its harmonics), the time-periodic part of the wavefunction can be expanded in the Fourier series,  $\Phi(t)=\sum_n e^{-in\omega t}\Phi_n$, and the index of the expansion harmonic $n$ can be interpreted as a position in the Floquet lattice. It has the physical meaning of the number of photons absorbed or emitted by the system.

The standard -- single-frequency -- Floquet theorem can be extended to the case of multiple incommensurate frequencies.  It has been utilized in the study of intense laser fields acting on atomic and molecular systems \cite{Ho83}, as well as to simulate the Anderson localization in dimensions higher than one \cite{Shep89}. In both cases, the physical system subjected to drive was zero-dimensional. The number of independent frequencies translates directly into the number of dimensions of the Floquet lattice. 
 
 \subsection{Wannier-Stark lattice and Berry curvature}

A peculiar feature of Floquet lattices generated by incommensurate frequency drives is that they  always experience an effective uniform ``electric field" applied in a non-crystallographic direction. The projection of the field onto a particular lattice direction is proportional to the corresponding drive frequency $\H_{\vec{\omega}}=\vec{n}\cdot \vec{\omega}$, where $\vec{n}$ is the integer vector of the drive harmonics, and $\vec{\omega}$ is a vector containing the angular frequencies of the drives.   By analogy with the one-dimensional Wannier-Stark ladders, if the potential energy drop over a lattice constant exceeds the hopping, then the band-structure effects are lost (this corresponds to the weak drive regime). On the other hand, for strong drive, the effective electric field (i.e., frequency) plays a role of a perturbation and the band description provides a convenient framework to study deviations from adiabaticity.

What are the physical implications of the topologically nontrivial Floquet band structure in the strong drive case? In conventional materials, bulk topological invariants, and Chern numbers in particular, lead to the appearance of gapless edge modes. In the case of Floquet lattice,  for a classical coherent drive, the lattice does not have a boundary. Nevertheless, the pseudo-electric field can induce chiral propagation, mimicking the edge physics in a finite crystal. Analogously, we can think of the effective force as inducing an anomalous velocity $\vec{\omega}\times\vec{\Omega}_{\vec{q}}$, where $\vec{\Omega}_{\vec{q}}$ is the Berry curvature in the Floquet Brillouin zone, defined below in Eq. (\ref{eq:BC}).  The chiral propagation on the Floquet lattice corresponds to the energy transfer between  individual drives (different frequencies). In the case of two-frequency drive of a two-level system, as we will show, the pumping power is proportional to the product of frequencies and the Chern number of the band into which system is initialized. Unlike the standard (perturbative) n-wave mixing, in the strong drive regime, any frequency can be converted into any other, no matter whether they are rationally or irrationally related.

\section{Floquet representation for multiple drive frequencies }\label{sec:2w}

\begin{figure}
\includegraphics[width=0.95\columnwidth]{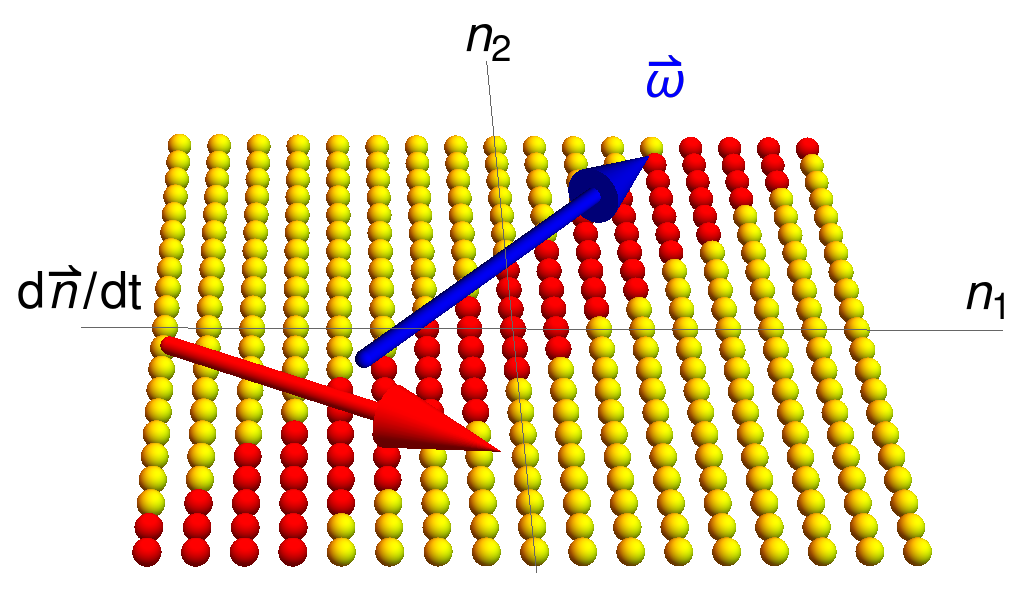}
\caption{The Floquet harmonics in a 2-drive system give rise to a 2-dimensional space. Each spot represents the $n_1\omega_1+n_2\omega_2=\vec{n}\cdot\vec{\omega}$ harmonic. If the frequencies are incommensurate, then the space is infinite. If the ratio $\omega_1/\omega_2$ is rational, however, the infinite lattice contracts to a cylinder  (strip marked in red, with periodic boundary conditions across the strip). The periodic drives appear to exert a force in the $\vec{n}$ space along the $\vec{\omega}$ direction. The motion that arises due to Berry curvature is normal to the force, and indicated by the $d\vec{n}/dt$ arrow. \label{rational}}
\end{figure}

The basis for our work is a mapping between a system of $d$ spatial dimensions subjected to $n$ mutually irrational drives, and a $d+n$-dimensional system. In rough terms, the extra $n$ dimensions in the multidrive Floquet problem emerge when we consider the number of energy quanta absorbed from each drive. As we show below, the numbers of photons absorbed from each drive make up the coordinates in an $n$-dimensional lattice. The energy associated with each photon, however, gives rise to an effective force in this lattice description, since the Hamiltonian will  contain diagonal terms $\sum_i n_i\omega_i$ with $n_i$ the number of absorbed photons from the drive with angular frequency $\omega_i$. Below we make this analogy more precise.

Consider a  system with basis states $|\alpha\rangle$, with $\alpha=1,2,\ldots$ (for instance, spin states, though  could also be position states in real space) subject to a Hamiltonian 
\be
\H=\summ_{\alpha,\beta} H^{\alpha\beta}(\varphi_1,\,\varphi_2,\ldots)\ket{\alpha}\bra{\beta},\label{h00}
\ee
such that each element $H^{\alpha\beta}(\varphi_1,\,\varphi_2,\ldots)$ is periodic for $\varphi_i\to\varphi_i+2\pi$. We assume linear time dependence for the $\varphi_i$'s
\be
\varphi_i(t)= \omega_i t,\label{eq:pht}
\ee
with the $\omega_i$'s being mutually incommensurate. The Schr\"odinger equation, written in terms of the components $\psi^\alpha$ of the wavefunction $|\psi\rangle = \sum_\alpha\psi^\alpha|\alpha\rangle$, is
\beq
i \partial_t \psi^\alpha(t) = H^{\alpha\beta}[\vec{\varphi}(t)]\psi^\beta(t).\label{eq:SE1}
\eeq

Eq. (\ref{eq:SE1}) represents a system under multi-tone drive. It can be analyzed in the spirit of Floquet theorem for a single drive. Following Floquet's construction, we write the wave function as
\beq
\ket{\psi(t)}=e^{-iEt}\summ_{\alpha,n_1,n_2,\ldots} \phi^\alpha_{\vec{n}} e^{-i\vec{n}\cdot\vec{\omega} t} \ket{\alpha}\label{eq:psi1}.
\eeq
Above we introduced the vector notation $\vec{n}\cdot\vec{\omega}=\summ_i n_i\omega_i$. The  Hamiltonian can also be expanded in terms of its Fourier components,
\be
\cH^{\alpha\beta}( \vec{\varphi}) = \sum_{\vec{p}}h^{\alpha\beta}_{\vec{p}}e^{ -i \vec{p}\cdot\vec{\varphi} }\label{htt}.
\ee
Next we combine Eqs. (\ref{eq:SE1}), (\ref{eq:psi1}) and (\ref{htt}) and obtain a tight-binding eigen-problem \footnote{it is important here that $\omega_1/\omega_2$ is irrational; for rational case see below}
\beq
(E + \vec{n}\cdot\vec{\omega})\phi^\alpha_{\vec{n}} = \summ_{\vec{p}} h^{\alpha\beta}_{\vec{p}} \phi^\beta_{\vec{n}-\vec{p}}.\label{eq:hop}
\eeq
Once more, we made use of the vector notation for the set of integers $p_i$.

Eq. (\ref{eq:hop}) describes a hopping problem on what we will refer to as a Floquet lattice (see Fig. \ref{rational}). It has as many dimensions as there are independent drive terms. In addition, it has a {\em tilt}, with  the potential $U(\vec{n}) = -\vec{n}\cdot\vec{\omega}$.  This makes intuitive the interpretation of the Floquet lattice: $n_i$ is the number of photons absorbed by the system from drive $i$. Accordingly, if the energy spectrum of the system in the absence of a drive is bounded, the wavefunction $\phi^\alpha_{\vec{n}}$ will be exponentially confined to a strip normal to the direction $\vec{\omega}$, that is, in the direction $(-\omega_2, \omega_1)$. Along the strips, the wavefunctions should be non-divergent, which fully specifies the eigenvalue-eigenvector problem $(\ref{eq:hop})$.  Naively, the number of eigenstates is equal to the lattice size times the dimension of the local Hilbert space. However, it is easy to see that the number of independent Floquet eigenstates is the same as for the undriven system. 
Indeed, equations (\ref{eq:hop}) have a symmetry: if $\phi^{(1)\alpha}_{\vec{n}}$ is a solution with quasienergy $E^{(1)}$, then  $\phi^{(2)\alpha}_{\vec{n}}=\phi^{(1)\alpha}_{\vec{n} - \vec{m}}$ is also a solution with $E^{(2)} = E^{(1)} -  \vec{m}\cdot\vec{\omega}$ for any integer vector  $\vec{m}$. Indeed, both correspond to the identical solution of the time-dependent Shr\"odinger equation Eq. (\ref{eq:SE1}). This can be verified explicitly, 
\bea
\psi^{(2)\alpha}(t) &=& \sum_{\vec{n}}e^{-i E^{(2)}t - i \vec{n}\cdot\vec{\omega} t}\phi^{(2)\alpha}_{\vec{n}} \nonumber\\
&=& \sum_{\vec{n}}e^{-i E^{(2)}t -  i \vec{n}\cdot\vec{\omega} t}\phi^{(1)\alpha}_{\vec{n}-\vec{m}}\nonumber\\
&=& \sum_{\vec{n}}e^{-i( E^{(2)} + \vec{m}\cdot\vec{\omega} )t - i  \vec{n}\cdot\vec{\omega} t}\phi^{(1)\alpha}_{\vec{n}}\nonumber\\
&=& \psi^{(1)\alpha}(t)\nonumber
\label{eq:psie}
\eea
Any initial wavefunction at $t=0$ can be expanded in terms of the unique Floquet eigenstates.

If it weren't for the linear onsite potential on the lhs of Eq. (\ref{eq:hop}), the problem would be a translationally invariant tight-binding model in the Floquet space and could be trivially solved by a Fourier transform. The Flqouet wave functions would be $\phi^{\alpha}_{\vec{n}}(\vec{\varphi})=\tilde{\phi}^{\alpha}(\vec{\varphi})e^{i\vec{\varphi}\cdot\vec{n}}$. Remarkably, the energy eigenvalues in the absence of the drive $\epsilon^{(i)}(\vec{\varphi})$ are the eigenvalues of $\H(\vec{\varphi})$ from Eq. (\ref{h00}). Thus, the angles $\vec{\varphi}$ play the role of momentum for the driven system. By analogy to the Bloch and Brillouin case, we refer to the region $0<\varphi_i<2\pi$ as the Floquet zone. 

With tilted potential on, the Floquet lattice eigenvalue problem becomes equivalent to Stark ladder in 2D \cite{Nakanishi93}. The tilt produces a change in the momentum-angles as
\be
\vec{\varphi}=\vec{\varphi}_0+\vec{\omega}t.\label{eq:pht}
\ee
To find the effects of the tilt on the wave functions, first consider the situation where the driving is strong, which is equivalent to the level spacings of $\epsilon^{(i)}(\vec{\varphi})$ being much greater than the largest angular frequency. In this case, we can describe the system using a semiclassical approach following the motion of a particle with momentum $\vec{\varphi}$ in the $\vec{n}$ lattice space. In this case, if the band structure  is topologically nontrivial, we can obtain chiral ``edge" modes that drift along the equi-potential lines in $\vec{n}$.  This will be central to our work. 

In the weak driving case, we obtain an effective 1D hopping model that describes quasicrystal with variable onsite potential (and variable hopping, if desired). This regime is not central to the main subject of the paper but interested readers may refer to Appendix \ref{App:weak}.

Let us next consider how the construction above changes when the frequencies have a rational ratio. Consider a 2-drive system, with $\omega_1 p= \omega_2 q$, with $p$ and $q$ mutually prime. First, the mapping above of the Floquet problem to a 2D lattice representing the different frequency components of the generalized Floquet wave function becomes redundant. The rational ratio can be taken into account by identifying the $(n_1,\,n_2)$ lattice point with $(n_1+mp,n_2-mq)$ for any integer $m$. This compactifies the 2D lattice into a strip with periodic boundary conditions, and the vector $(p,\,-q)$ connects equivalent points across the strip's boundary. So, essentially, the problem is thereby reduced to that of a cylinder of circumference $\sqrt{p^2+ q^2}$, made of a square lattice (Fig. \ref{rational}; for detailed treatment see Appendix \ref{sec:cd}). 
As we will show in the following, for large $\sqrt{p^2+ q^2}$ there is no significant difference between the rational and irrational cases as far the energy pumping efficiency is concerned. This is indeed to be expected since for a local Hamiltonian (not many higher harmonics) the system is only sensitive to the local geometry (same for plane and cylinder), and not the global topology.

A related observation, interesting from mathematical perspective, is that it is possible to define Floquet eigenstates for quasiperiodic drive, by means of a limiting procedure.
Specifically, we can approximate any irrational frequency ratio as a limit of a ratio of two integers both tending to infinity, $\omega_1/\omega_2=\mbox{lim}_{i \rightarrow\infty}\frac{q_i}{p_i}$. Solving a sequence of Floquet eigenstate problems  for progressively smaller $\omega_i=\frac{\omega_1}{q_i}=\frac{\omega_2}{p_i}$ we find that the eigenstates indeed converge to a limiting state that one can define as the quasiperiodic Floquet eigenstate. The demonstration of this result is given in Appendix \ref{App:floq}.

\section{Temporal Topological systems}\label{sec:BHZ}

\begin{figure}
\includegraphics[width=0.9\columnwidth]{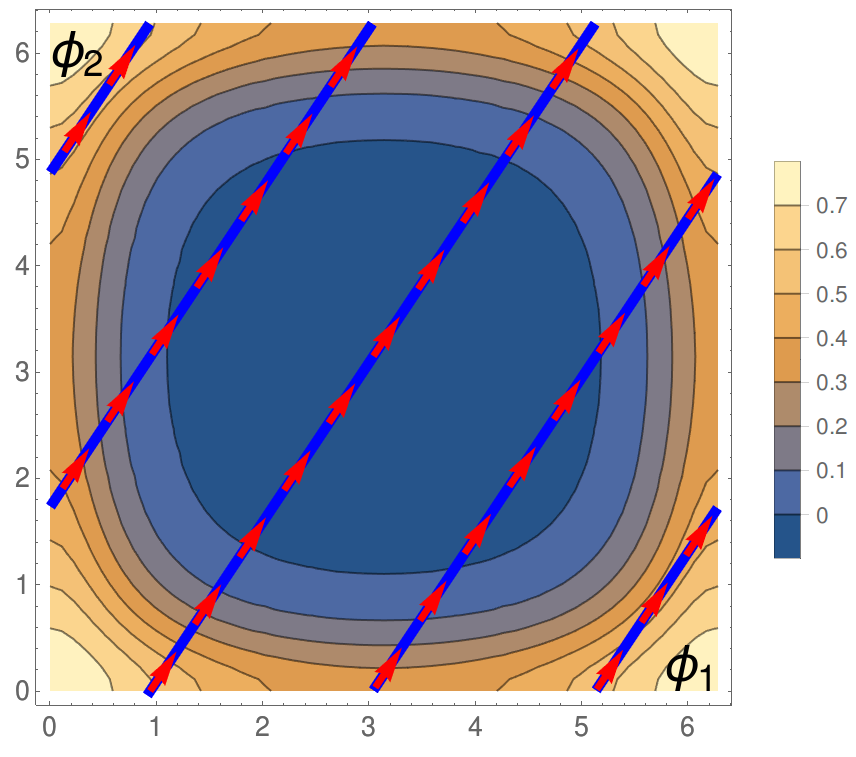}
\caption{The Floquet zone for two drives. In analogy with a 2D band structure, we draw the Berry curvature (color plot with side legend) vs. the two offset phases $\varphi_1$ and $\varphi_2$. The periodic drives are akin to a motion along this Floquet zone with $\vec{\varphi}=\vec{\varphi}_0+\vec{\omega} t$. In the case of a rational frequency ratio, the system will explore a closed periodic path through the Floquet zone. Drawn here is an example of $3\omega_1=2\omega_2$. The pumping effect will be roughly the integral of the Berry curvature along the path times $\omega_1\omega_2$ (Eq. \ref{eq:P}). \label{FZpath}}
\end{figure}

The analogy between multiple incommensurate drives and multi-dimensional lattice quantum dynamics allows us to constructs 0-dimensional topological systems. The topological effects will arise from the temporal structure of the wave functions of the driven system. Below we construct such a system which consists of a single spin-1/2, driven by two incommensurate periodic drives. We first define the model, and then explore its topological properties. Particularly, we will consider the semiclassical motion of the system on the Floquet lattice. We will concentrate on the associated $\vec{\varphi}$ momentum space, especially when there is Berry-curvature associated with the Floquet-lattice momentum states $\psi^{\alpha}(\vec{\varphi})$. Fig. \ref{FZpath} depicts all elements relevant to the discussion, and maps out the motion of the system in the momentum $\vec{\varphi}$ space, as well as the Berry curvature of the model defined below. 

\subsection{ The BHZ path to temporal topological physics}

 Let us choose $h^{\alpha\beta}_{pq}$  in Eq. (\ref{eq:hop}) so that the translationally invariant tight binding band structure is topologically non-trivial. One of the simplest band structures of this type is a half of the BHZ model \cite{Bernevig06}, with the Hamiltonian

\bea
\H=&v_x \sin(k_1)\sx+v_y \sin(k_2)\sy\nonumber\\
&+\l[m-b_x\cos(k_1)-b_y\cos(k_2)\rr]\sz.\label{2f}
\eea

The model yields a quantum Hall insulator (band Chern numbers $\pm 1$) for $-|b_1| -|b_2|<m< -\left||b_1| -|b_2|\right|$ and $\left||b_1| -|b_2|\right|<m< |b_1| +|b_2|$.
The corresponding Floquet Hamiltonian is obtained by replacing $k_1\to \omega_1 t+\varphi_1$ and $k_2\to \omega_2 t+\varphi_2$,
\bea
\H&=v_1 \sin(\omega_1 t+\varphi_1)\sx+v_2\sin(\omega_2 t+\varphi_2)\sy \nonumber\\
& +\l[m-b_1\cos(\omega_1t+\varphi_1)-b_2\cos(\omega_2t+\varphi_2)\rr]\sz.\label{eq:2f}
\eea
The interpretation of this Hamiltonian harks back to the discussion of the emerging Floquet lattice of Sec. \ref{sec:2w}. 
The operator $e^{-i\omega_i t-i\varphi_i}$ absorbs a photon from drive $i$, thus realizing a hop on the Floquet lattice in the direction $i$. In order to be able to follow the time evolution on the Floquet lattice, let us explicitly extend the Hilbert space to the direct product of the spin and lattice spaces, with the wavefunction $\ket{\psi} = \sum_{\alpha\vec{n}} \psi^\alpha_{\vec{n}}\ket{\vec{n}}\ket{\alpha}$ [compare with Eq. (\ref{eq:psi1})]. In this representation, $e^{\pm i\omega_i t} \to \ket{n_i\mp1}\bra{n_i}$.
We obtain a tight binding Schr\"odinger equation for the wave function amplitude on the Floquet lattice 
\begin{widetext}
 \be
 \ba{c}
i\partial_t\psi_{n_1,n_2} =  \H\psi |_{n_1,n_2}=
 \frac{1}{2}\l(i v_1\sx-b_1\sz\rr)  e^{i\varphi_1}\psi_{n_1-1,n_2}+\frac{1}{2}\l(-i v_1\sx-b_1\sz\rr)  e^{-i\varphi_1}\psi_{n_1+1,n_2} \vspace{2mm}\\
 +\frac{1}{2}\l(i v_2\sy-b_2\sz\rr)e^{i\varphi_2}\psi_{n_1,n_2-1}+ \frac{1}{2}\l(-i v_2\sy-b_2\sz\rr)e^{-i\varphi_2}\psi_{n_1,n_2+1}\vspace{2mm}\\
 +\l(m\sz - n_1\omega_1-n_2\omega_2\rr)\psi_{n_1,n_2}
 \ea
 \ee
We could even go further and Fourier transform the hopping part of the Hamiltonian, to obtain its `momentum representation' on the tight-binding Floquet lattice:
 \be\ba{c}
 \H_{\vec{q}}=v_1\sx \sin(q_1+\varphi_1)+v_2\sy \sin(q_2 +\varphi_2)
 +\l[m-b_1\cos(q_1+\varphi_1)-b_2\cos(q_2+\varphi_2)\rr]\sz\vspace{2mm}\\
 \H_{\vec{n}}=-\vec{n}\cdot\vec{\omega}\vspace{2mm}\\
 \H=\summ_{\vec{q}}\H_{\vec{q}}\hat{n}_{\vec{q}}+\summ_{\vec{n}}\H_{\vec{n}}\hat{n}_n
 \label{2fFloquet}
 \ea
 \ee
 where $\hat{n}_{\vec{n}}$ is the occupation of site $\vec{n}=(n_1,n_2)$ and similarly  $\hat{n}_{\vec{q}}$ is the occupation of the momentum state $(q_1,q_2)=\vec{q}$. Hamiltonian written in the form Eq. (\ref{2fFloquet}) mixes the real (frequency) and momentum (phase/time) space representations. The diagonal part $\H_{\vec{n}}=\vec{n}\cdot\vec{\omega}$  has the role of a force, which pushes the momenta $\vec{q}$ in the $\vec{\omega}$ direction. This naturally leads to the momentum evolution
 \be
 \vec{q}=\vec{\omega} t.
 \ee
 
\end{widetext}

\subsection{Photon absorption rates}\label{sec:pump}

What would be the manifestation of a topological band structure in the Floquet lattice? When the Berry curvature is present, the system will respond to the force implied by $\H_{\vec{n}}=-\vec{n}\cdot\vec{\omega}$ by moving normal to the force in a preferred direction determined by the band into which the system was initiated.
From the practical stand point, the ``movement" of the particle within a chiral band corresponds to a process where photons of one frequency are absorbed and of the other are emitted, in the proportions that $approximately$ (up to uncertainty of the scale $||h(t)||$) conserve energy. In other words, the rates of work performed by ``sources'' with frequencies $\omega_i$ should add up to approximately zero on average, but each one can be substantial.  

Let us derive expressions for the photon emission and absorption. From Eq. (\ref{2fFloquet}) we readily notice that the offset phases $\varphi_i$ act as a lattice vector potential for the photon flow. Therefore, the derivative
\be
\hat{j}_i=\der{\H}{\varphi_i}=\der{n_i}{t}\label{jx}
\ee
is the current, or velocity operator, in the Floquet lattice. It is the rate of absorption/emission of photons with frequency $\omega_i$.
The rate of energy absorption (power) is then
\be
\der{\langle E_i\rangle}{t}=\omega_i \langle \hat{j}_i\rangle =\omega_i\der{\langle \hat{n}_i\rangle}{t}.
\ee
This intuitive result can be also obtained in another way. Consider a generic double drive Hamiltonian as $\H=\vec{h}_1(\omega_1t+\varphi_1)\cdot\vec{\sigma}+\vec{h}_2(\omega_2t+\varphi_2)\cdot\vec{\sigma}$. The time-derivative of the total energy is $dE/dt =d\braket{H}/dt=i\braket{[H,H]} + \braket{\der{H}{t}} = \der{\vec{h}_1(\omega_1t+\varphi_1)}{t}\cdot\braket{\vec{\sigma}}+\der{\vec{h}_2(\omega_2t+\varphi_2)}{t}\cdot\braket{\vec{\sigma}}$.
Therefore, change of energy due to a given source is 
\be
\der{\langle E_i\rangle}{t}=\omega_i\l\langle \der{\H}{\phi_i}\rr\rangle=\der{\vec{h}_i}{t}\cdot{\langle\vec{\sigma}\rangle}.\label{eq:P}
\ee
Note that the same formalism can be applied to determine not only the work performed by a given frequency drive, but even to separate the energy flows between different polarizations of the same frequency drive. 

 \subsection{Semiclassical equations of motion and pumping power}
 
 The evolution of position on the Floquet lattice in the limit of small applied ``electric" field (that is, small frequency) can be determined from the semiclassical equations of motion. Within a particular band of $\H_{\vec{q}}$, they are \cite{Niu99},
 \be
\ba{c}
 \dot{\vec{n}}=\nabla_{\vec{q}}\epsilon_{\vec{q}}-\nabla_n\H_{\vec{n}}\times\Omega_{\vec{q}}\vspace{2mm}\\
 \dot{\vec{q}}=-\nabla_n\H_{\vec{n}}=\vec{\omega}.\label{eq:SC}
 \ea
 \ee
In the Floquet problem, there is a constant ``force", $\vec{\omega}$, and we  take
 \be
 \vec{q}(t)=\vec{\omega}t,
 \ee
which after substitution into Eq. (\ref{2fFloquet}) -- as expected -- yields the original problem, Eq. (\ref{eq:2f}). The key to our problem is the anomalous velocity, related to the Berry curvature
 \be
 \Omega_{\vec{q}}=\hat{z} i\frac{1}{2}\mbox{tr}\l(P_{\vec{q}}\l[\der{P_{\vec{q}}}{q_1},\der{P_{\vec{q}}}{q_2}\rr]\rr)\label{eq:BC}
 \ee
 with $P_{\vec{q}}$ a projector onto a particular band of  $\H_{\vec{q}}$ [e.g. for the lower band, $P_{\vec{q}}=\l(1-{\H_{\vec{q}}}/{\epsilon_{\vec{q}}}\rr)/2$].

 On average, only the Berry curvature pumps. The energy that goes between the two drives is then
 \be
\frac{1}{2}\der{(E_1-E_2)}{t}=\frac{1}{2}(\omega_1,-\omega_2)\cdot\l(\vec{\omega}\times\Omega_{\vec{q}}\rr)=|\Omega_{\vec{q}}|\omega_1\omega_2\label{eq:P12}.
 \ee

\subsection{Quantum Hall analogy}

The above result can also be obtained  by directly exploiting the analogy with the Hall response in the topological insulators. Namely, we will use the known expression for the Hall current in order to determine the average drift velocity of an individual particle. 
We take unit cell size 1 and electric field $\vec{\cal E} = (\omega_1, \omega_2)$. The field is pointing at angle $\alpha = \arctan\omega_2/\omega_1$ to  axis 1. For one fully occupied band with Hall conductivity $\sigma_{xy}$, the application of the electric field leads to the  transverse current density
$$  j_\perp = \sigma_{xy} {\cal E}$$
that flows in the direction $\alpha + \pi/2$. This current is a product of particle density and velocity, and since the size of a unit cell is one, and the band is fully occupied, the density is 1.  Thus, the drift velocity is $v = \sigma_{xy} \sqrt{\omega_1^2 + \omega_2^2}$.  Its projection onto axis 1 is $v_1 = v \sin\alpha = \sigma_{xy}\omega_2$, and, therefore, the rate of energy change along the axis 1, which is the same as the power exerted by mode 1 is 
$$\der{E_1}{t} = \sigma_{xy}\omega_1\omega_2 = -\der{E_2}{t},$$
equivalent to Eq. (\ref{eq:P12}).  There is also an appealing connection to quantum pumping, described in Appendix \ref{sec:QP}.

 \subsection{Role of incommensurability} \label{sec:com}
 
 If $\omega_1/\omega_2$ is a rational number, then, over time, the path in the Floquet zone $\vec{\omega} t \, \mbox{mod}\, 2\pi$ will repeat itself, and only a part of the Berry curvature  will be sampled. On the other hand, an irrational ratio would sample uniformly the entire double Floquet zone. Therefore, for rational  $\omega_1/\omega_2$ the pumping power is not generally quantized; however,  the average over the phases $\varphi_1$ and $\varphi_2$ (which can be interpreted as initial conditions) is. 

\section{Results of numerical simulations\label{nums}}

Let us now explore the pumping effects described above numerically. We find that, indeed, in the topological parameter range, as long as the gap in the corresponding band Hamiltonian exceeds the drive frequencies, i.e., for sufficiently strong drive, energy flows between the two drives at a nearly quantized rate. A necessary condition for strong pumping is the high `fidelity' -- large projection of the spin onto the direction of the instantaneous `magnetic field' (equivalent to the ability of the spin to stay  in one topological band).
We consider these for both incommensurate  and commensurate drive frequencies. Interestingly, depending on the initial conditions, the pumping effect for rational frequency ratio can exceed its counterpart for incommensurate systems, and persist even outside the topological regime.  The qualitative reason lies in the incomplete sampling of the Floquet zone in the case of rational drive, and thus a possibility of preferential sampling of high Berry curvature regions.

\subsection{Technical interlude\label{tech}}

Before diving into the numerical results, several technical aspects of the simulation must be discussed. First, in our simulation we integrated the Schr\"odinger equation to produce the unitary evolution operator $U(t)$.  The Hamiltonian we use is a special case of Eq. (\ref{eq:2f}):
\bea
\frac{\H}{\eta}&=\sin(\omega_1 t+\phi_1)\sx+\sin(\omega_2 t+\phi_2)\sy \nonumber\\
& +\l[m-\cos(\omega_1t+\phi_1)-\cos(\omega_2t+\phi_2)\rr]\sz\label{eq:2fs},
\eea
with the parameter $\eta$ defining the overall energy scale. Roughly, the ratio of $\eta$ to the fastest frequency $\omega_i$ determines whether the strong or weak driving regime is realized. 
For the gap parameter $m$ we consider the topological ($|m| < 2$) and non-topological ($|m| > 2$) regimes.
In our simulation, we constructed $U(t)$ as a product of the `infinitesimal' evolution matrices $\exp\l[-i\H(t)dt\rr]$. Our time discretization step was $dt=0.001$, such that $\omega_1 dt=10^{-4}$. 

Given the unitary evolution operator, it is possible to calculate the integrated work done by each of the two drives. As discussed in Sec. \ref{sec:pump}, the instantaneous power spent or absorbed by drive $i$ is 
\be
\frac{dW_i}{dt}=\bra{\psi(t)}\frac{dh_i(t)}{dt}\ket{\psi(t)}=\bra{\psi_0}U(t)^{\dagger}\frac{dh_i(t)}{dt}U(t)\ket{\psi_0}
\ee
[Here, $h_i(t) = \vec {h}_i(t)\cdot \vec{\sigma}$]. Therefore we can define the work operator:
\be
\hat{W}_i=\intt_0^tdt U(t)^{\dagger}\frac{dh_i(t)}{dt}U(t),
\ee
such that $W_i=\bra{\psi_0}\hat{W}_i\ket{\psi_0}$. Alongside $U(t)$, we also calculated the operators $\hat{W}_i$ for the two drives. 

Finally, we must discuss the initial conditions for the simulation. We would like to have an initial state $\ket{\psi_0}$ that would maximize the energy transfer between the two drives in the  topological case. For this purpose we hark back to the analogy between the time-dependent Hamiltonian [e.g., as written in Eq. (\ref{eq:2fs})], and the Hamiltonian of a real-space topological insulator. The analogy, and the semiclassical logic of our topological pumping arguments suggest that a good choice for the initial state is an eigenstate of the instantaneous Hamiltonian at the beginning of the drive. When the drive is strong, $\eta\gg \omega_{1,2}$, the motion of the system in the photon-number space will be dictated by the semiclassical equations (\ref{eq:SC}), with the Berry curvature determined by the band of the eigenstate $\ket{\psi_0}$. In this strong-drive limit, the eigenstates of the instantaneous initial Hamiltonian coincide with the Floquet eigenstates of a periodic system.  For weaker drives, it might be advantageous to initialize the system into a Floquet state of a periodic system which approximates the incommensurate two-drive system.  We explore this briefly in Appendix \ref{sec:FvsI}.

The fidelity of the initial state evolution, $\psi(t) = U(t)\psi_0$ with respect to the same-band instantaneous eigenstate of $\H(t)$, $\psi_i(t)$, 
 is crucial for effective driving in the incommensurate case. It is possible to write the fidelity as
\bea
\cF &=& \bra{\psi(t)}P_t\ket{\psi(t)}\nonumber\\
&=&\mbox{tr}\l[\l(\frac{1}{2}-\frac{\H(0)}{2\epsilon(0)}\rr)U^{\dagger}(t)\l(\frac{1}{2}-\frac{\H(t)}{2\epsilon(t)}\rr)U(t)\rr].\label{eq:fid}
\eea
From the definition of the projection operator at time $t$, $P_t$, it follows that 
\be
\cF = \frac12\left[1+ \braket{\vec\sigma(t)}\cdot \frac{\vec h(t)}{h(t)}\right] \equiv \cos^2 \frac{\theta(t)}{2}, 
\ee
where $\theta(t)$ is the angle between the spin expectation and the instantaneous magnetic field at time $t$.
If the spin is not in an instantaneous eigenstate, then the semiclassical equations of motion do not apply, and the pumping effect is suppressed.  The fidelity is robust, however, as long as the minimum gap in the band Hamiltonian significantly exceeds the drive frequencies,
\be
\Delta=\eta\min(||m|-2|, |m|)\gg\omega_1,\,\omega_2.
\ee
 
When studying the rational frequency ratio case, it is natural to consider the Floquet eigenstates as initial state for the spin in addition to the instantaneous eigenstates.  A rational frequency ratio implies a strict periodicity of the Hamiltonian. For $\omega_1=2\pi/Tp$ and $\omega_2=2\pi/Tq$, the Hamiltonian is periodic with period $\tau=T \mbox{LCM}(p,q)$, with $\mbox{LCM}(p,q)$ the lowest common multiple of $p$ and $q$. Typically, initialization into Floquet eigenstates results in faster pumping at short and intermediate times, but more importantly, since they are the eigenstates of $U({\tau})$, the pumping power itself is a periodic function of time.

\subsection{Incommensurate drives}

\begin{figure}
a.\includegraphics[width=.95\columnwidth]{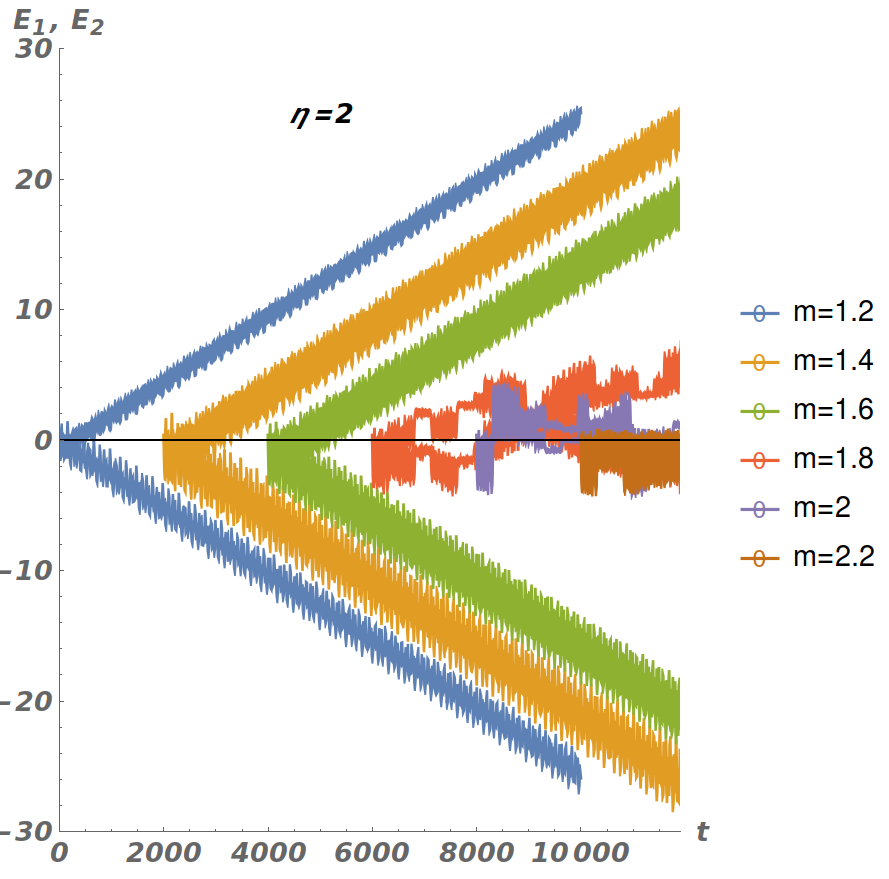}

b.\includegraphics[width=.95\columnwidth]{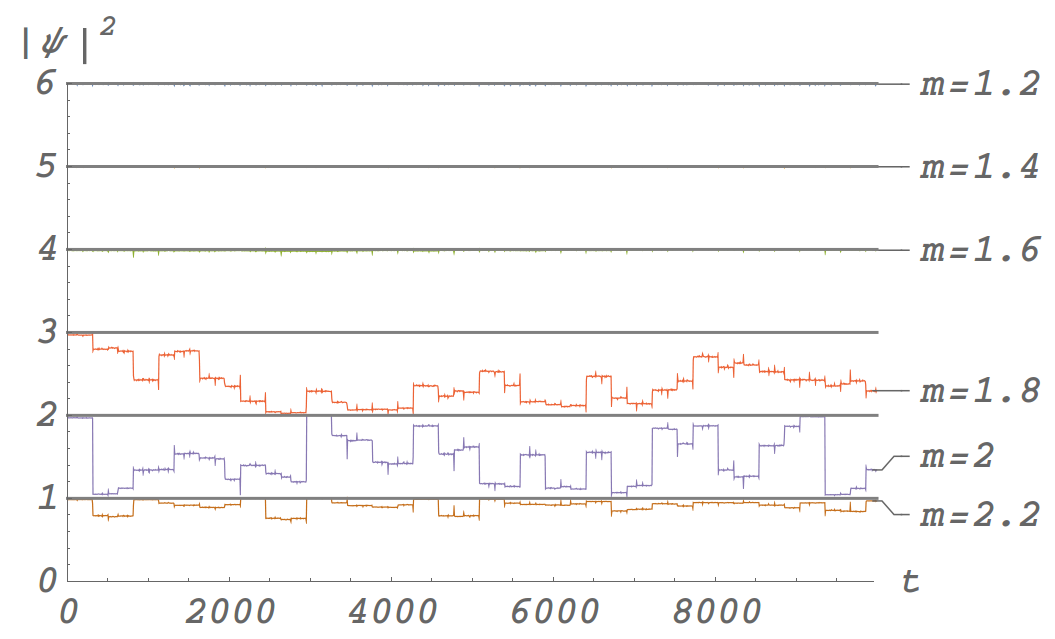}

c.\includegraphics[width=.95\columnwidth]{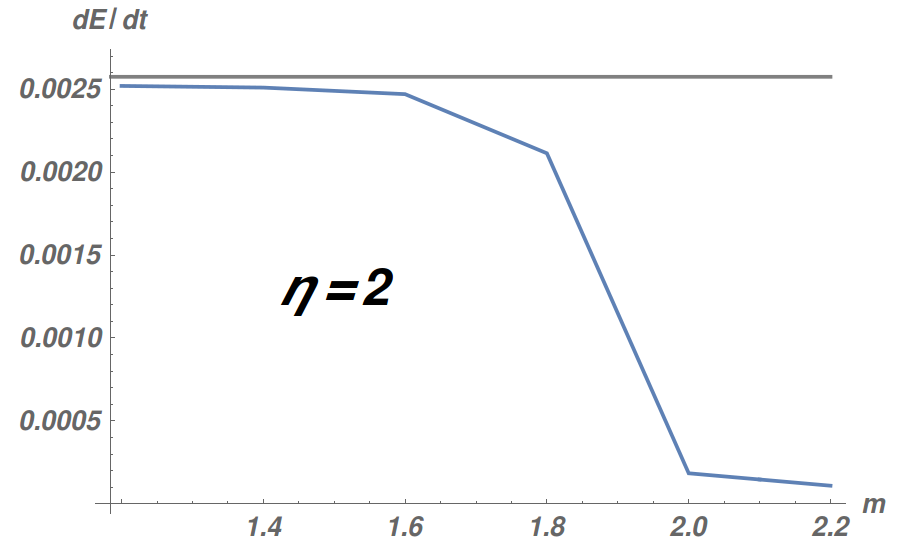}

\caption{The energy flow and fidelity in the strong-drive regime, $\eta=2$. (a) The total work done by drives 1 and 2 as a function of time for different $m$'s. Each pair of lines is displaced on the time axis for clarity. (b) Fidelity vs. time of a state initialized into the instantaneous eigenstate of $\H(t=0)$, and measured against the eigenstate of $\H(t)$. (c) The power pumped as a function of the parameter $m$ averaged up to time $t=2000$. The quantized pumping is marked by the gay line at $dE/dt=\omega_1\omega_2/2\pi$.  The strong-drive regime realizes the topological pumping prediction up to $m=1.8$, with the energy pumping rate saturating near the expected value of $dE/dt=\frac{\omega_1\omega_2}{2\pi}$ for a model with Chern-number 1. The initial phases for these plots are $\phi_1=\pi/10,\,\phi_2=0$. We see that the fidelity of the evolved state relative to the single-band state deteriorates as we get close to the phase transition. This is due to the closing of the band gap in the corresponding BHZ model. \label{strong-flow}}
\end{figure}

Let us demonstrate the pumping effects in the regime that reflects best the semiclassical limit, in which the Berry curvature effects are dominant. Simulating the system at $\eta=2$, yields near perfect pumping effects all the way essentially to the topological transition at $m=2$. In Fig. \ref{strong-flow} we see the energy transfer for several values of the gap parameter $m$, alongside the fidelity associated with the overlap of the initial state with the instantaneous  eigenstate [as defined in Eq. (\ref{eq:fid})]. 

Throughout our analysis here, we use the incommensurate frequency pair $\omega_1=0.1$ and $\omega_2=\gamma\omega_1$, with $\gamma=\frac{1}{2}\l(\sqrt{5}+1\rr)\approx1.618$, the golden ratio. The offset phases are $\phi_1=\pi/10,\phi_2=0$. We will explore the evolution up to times $t=10^4$, where the pumping effect is clearly visible. We will keep the parameters $v=1$ and $b=1$ fixed, and vary $m$ and $\eta$ in our exploration.

The energy pumping rate indeed saturates to the one expected from the semi-classical analysis. For a system with Chern number 1 we expect an average Berry curvature of $\Omega=\frac{1}{2\pi}$, and an energy pumping rate of $dE/dt=\frac{\omega_1\omega_2}{2\pi}=0.00257518$, as in Eq. (\ref{eq:P12}). From linear-regression of the plots in Fig. \ref{strong-flow} we obtain excellent agreement with the semiclassical theory. Similarly, the fidelity also remains near perfect  throughout the evolution, except for  the case of $m=1.8$. There, $\Delta=\eta|m-2b|=0.4$, which is comparable to $\omega_2\approx 0.1618$, and the quasi-adiabaticity condition is expected to break down. 

\begin{figure}
\includegraphics[width=0.95\columnwidth]{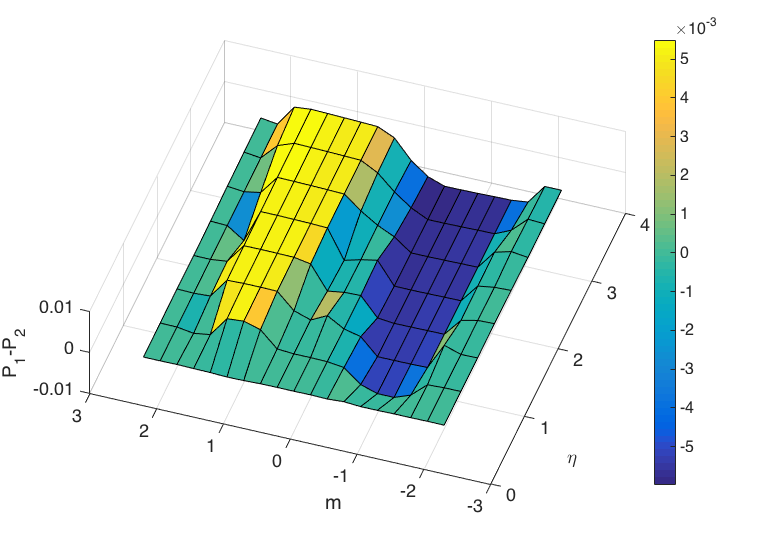}
\caption{Pumping rates for different $\eta$ and $m$. The parameters of the system for this plot are $b=v=1$, and frequencies $\omega_1=0.1$, and $\omega_2=\gamma \omega_1$. The pumping rate was averaged over the time range $t<10000$. \label{Wpd}}
\end{figure}

As $\eta$ is reduced, the regime of ineffective pumping, where the gap $\Delta$ becomes comparable to or smaller than $\omega_{1,2}$, expands. This is visible in Fig. \ref{Wpd}, which shows the pumping rate as a function of $m$ and $\eta$ for a broad region of parameter space. We explore smaller values of $\eta$, as well as the demise of this effect at weaker drives in Appendix \ref{sec:moreNums}.

\subsection{Commensurate drive frequencies\label{sec:rat}}

The energy pumping effect, and Eq. (\ref{eq:P}) in particular, can also be explored for a rational frequency ratio. In this case, however, it is not the Chern number that determines the effect, but rather the integral of the Berry curvature along a closed path in the two-dimensional `Floquet zone' , as shown in Fig. \ref{FZpath}.  We study this regime here for the frequency ratio $\omega_1/\omega_2=2/3$ (with $\omega_1=0.1$) and the strong drive, $\eta=2$. Additional results for weaker drives are given in the Appendix \ref{sec:moreNums}.

As discussed in Sec. \ref{tech}, when the frequency ratio is rational, the Hamiltonian is strictly periodic, and we can initiate the spin into a Floquet eigenstate of $\H(t)$. This  results in a periodic pumping profile and typically larger pumping power, compared to the initialization into instantaneous eigenstates (detailed comparison between the two initiation procedures is given in Appendix \ref{sec:FvsI}). 

\begin{figure}
a.\includegraphics[width=.85\columnwidth]{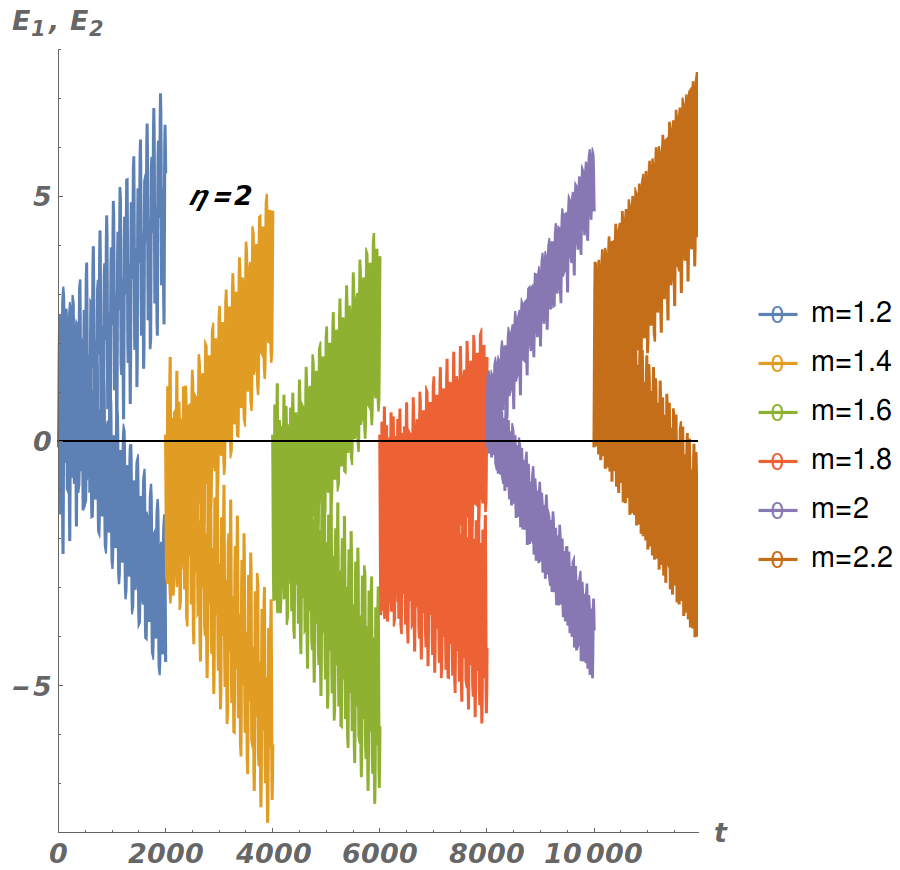}

b.\includegraphics[width=.85\columnwidth]{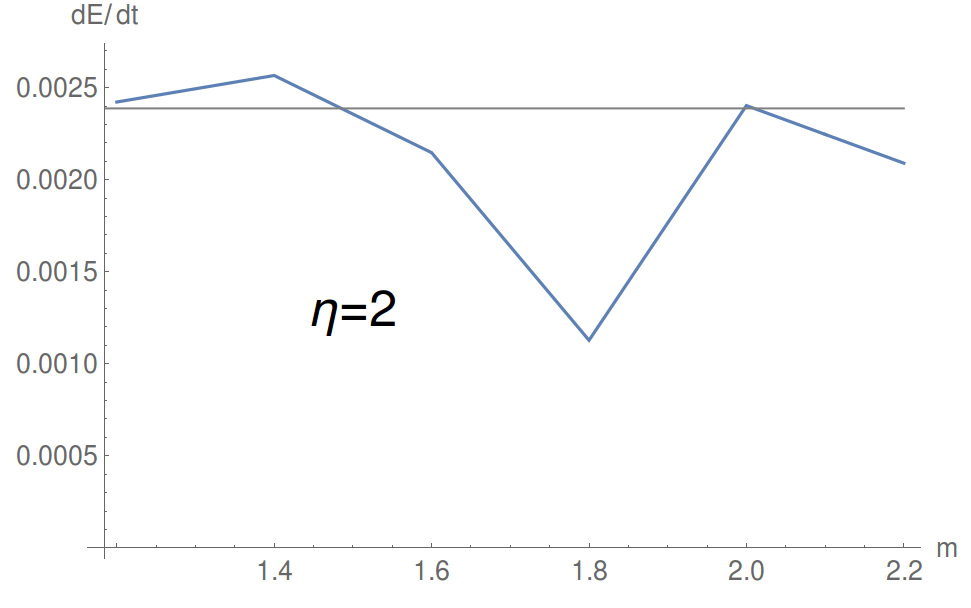}
\caption{The energy flow in the strong-drive regime, $\eta=2$, for rational frequency ratio, $\omega_2=\frac{3}{2}\omega_1$, with Floquet state initialization. (a) The total work done by drives 1 and 2 as a function of time for different $m$'s. Each pair of lines is displaced on the time axis for clarity.  (b) The power pumped as a function of the parameter $m$ averaged upto time $t=2000$. The pumping in the commensurate case could be stronger than its incommensurate counterpart. Furthermore, pumping may persist in the non-topological regime, since the system does not average the Berry curvature of the entire Floquet zone. In the case of the offset angles chosen, even when $m=2.2$, the system explore regions with net  positive Berry curvature. The gray line is at $dE/dt=\omega_1\omega_2/2\pi$, signifying the pumping due to Berry curvature of Chern number $C=1$ band. The initial phases for these plots are $\phi_1=\pi/10,\,\phi_2=0$.  \label{strFW}}
\end{figure}

\begin{figure}
a.\includegraphics[width=0.85\columnwidth]{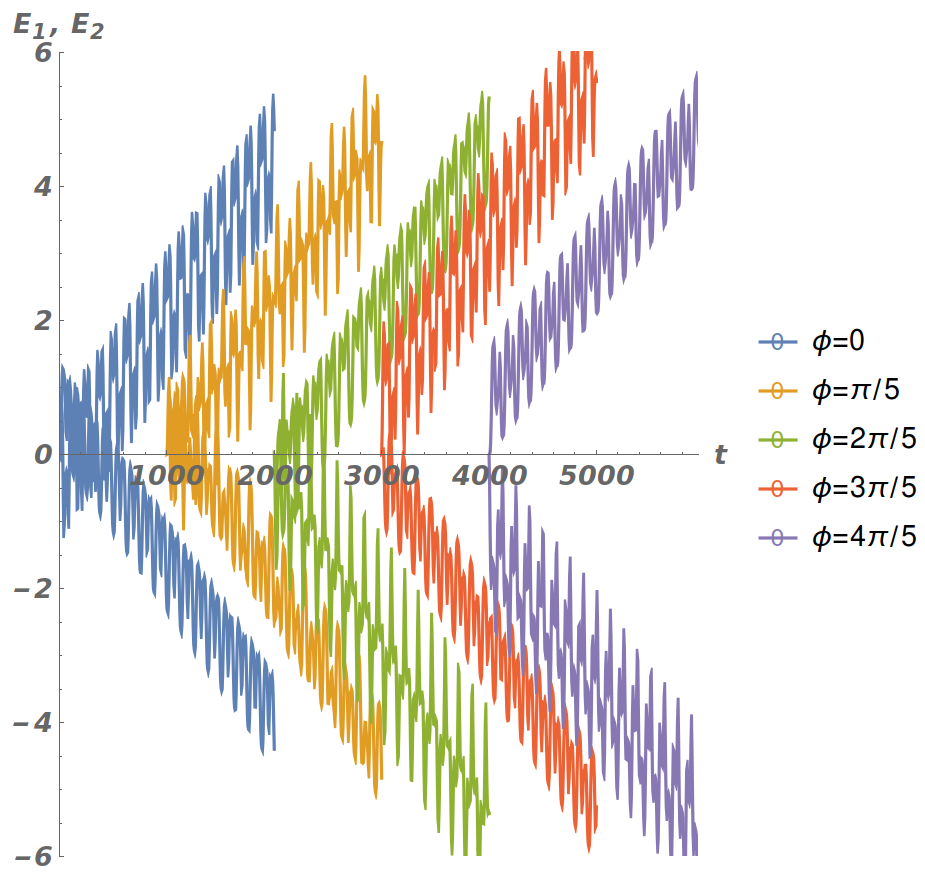}

b.\includegraphics[width=0.85\columnwidth]{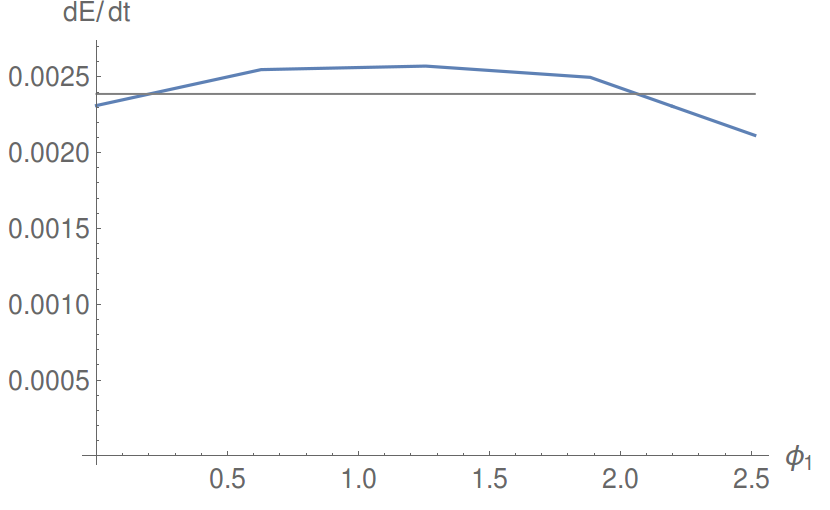}
\caption{Pumping profile for different offset phases: (a) Accumulated work for Floquet initial states, (b) average power for $t<2000$. We take $m=b=v=1$,  $\eta=1$, and  $\phi_1=n\pi/5$ with $\frac{5}{\pi}\phi_1=0,1,2,3,4$. The Floquet zone path of this system for one offset phase combination is shown is Fig. \ref{FZpath}). Floquet eigenstates initial conditions were used. For these parameters there were no significant differences with instantaneous eigenstate initialization.  \label{Wphi}}
\end{figure}

The pumping profile for $1.2<m<2.2$, with offset phases $\phi_1=\pi/10,\,\phi_2=0$,  and Floquet eigenstate initialization is shown in Fig. \ref{strFW}. As can be seen, the energy pumping effect is present in the entire topological region. Indeed, the pumping rate may exceed the quantized theoretical value, since the energy pumping effect is now determined by the Berry curvature along a particular periodic path through the Floquet zone, rather than its average over the whole Floquet zone.
Different paths can be selected by varying the offset phases. This dependence is illustrated in Fig. \ref{Wphi}. It is also interesting to note that the offset-phase dependence is stronger for initialization into Floquet eigenstates, rather than into instantaneous eigenstates of the Hamiltonian (not shown).

\section{Summary and discussion \label{disc}}

In this work we have demonstrated that it is possible to implement multidimensional topological band structures purely in the frequency space. The implementation relies on the ability to apply strong drives of incommensurate frequencies. The 
approach was illustrated using the 2-dimensional BHZ \cite{Bernevig06} model constructed by pumping a two-level system (``spin-1/2") by two elliptically polarized waves of incommensurate frequencies. The observable in this case is the quantized energy pumping power between the two  drives, which is the direct analog of the quantized transverse Hall conductivity of the original real-space model. As such, the driving occurs in the topological regime of the model, and remains effective as long as the drives' frequencies are lower than the minimal band gap in the model (quasi-adiabatic regime).

There are two ways in which the model can be naturally generalized: by considering larger spins, and by allowing the driving fields to have their own dynamics. The generalization of our results to larger spins is straightforward. Indeed, the dynamics of the expectation value of the spin is given by the Bloch equations, which are independent of the spin size \cite {abragam1961principles}, 
\bea
\dot{\vec S}(t) = i[H, \vec S(t)] =\vec S(t)\times  {\vec h}(t)\label{eq:Bloch}.
\eea
The Bloch equations are linear in spin, and therefore the precession frequency is independent of the spin size.
The ``strong drive" criterion needed for the energy pumping for general  spin size, in the language of the Bloch equations, corresponds to having the instantaneous Larmor frequency $|\vec h(t)|$ higher than the pump frequencies $\omega_i$ --  same condition as for spin-1/2.  On the other hand, from Eq. (\ref{eq:P}), the pumping power depends linearly on the spin size, and thus by taking a large magnetic particle with low magnetic anisotropy (e.g. YIG sphere \cite{YIG}), or NMR, or ESR systems \cite{Si:P} one can dramatically increase the pumping power, to macroscopic levels.

One practical way to include the dynamics of the drive fields is by considering a closed system of spin and two electromagnetic cavities resonating at the drive frequencies. The cavities can be initialized into semiclassical coherent states to represent the periodic drives. The energy pumping will act to change both amplitudes (and phases) of the drives, eventually taking the system out of the topological regime.  Our preliminary results indeed show that system reaches the topological state boundary, after which the pumping direction reverses, albeit at a reduce rate.

Treating photons quantum mechanically raises several new questions. Perhaps most appealing is the possibility of using the two-frequency topological effects to pump energy from a laser of one frequency into a cavity with another frequency. Additionally: \\
-- What kinds of entangled spin-photon states are achievable?\\
-- Is there is a spontaneous transition into the pumping regime if one of the cavities is initialized into a vacuum state?\\
--  Can a superradiance transition occur in a system of multiple spins in the cavities? \\
We leave these theoretical problems for future study.

The quantum limit may be particularly amenable to experimental study using superconducting qubit systems \cite{Manucharyan113}.  Their appeal of this scheme is both in the exquisite level of control over the qubit, as well as in the extreme measurement sensitivity to the photon occupation numbers in the superconducting resonant circuitry.
Yet another attractive feature of the superconducting devices, and Josephson junctions in particular, is the access to the AC Josephson effect in order to drive the superconducting phase by applying DC voltage. 
Manipulation of the superconducting phases has indeed been considered as a means to explore and control topological phases in multi-terminal Josephson junctions \cite{Heck, Meyer}, and is mathematically analogous to the phase control provided by the AC drive that we consider in the present work.

In this work we considered only perfectly stable (fixed-frequency) drive sources. We have seen, however, that there is no significant difference between commensurate and incommensurate drives of approximately equal frequencies, as far as the pumping properties are concerned.  It would be interesting to consider more general phase dynamics, deviating from the one given by Eq. (\ref{eq:pht}), either due to the deliberate tuning of drive frequencies,  or due to noise.

\section{Outlook}\label{sec:out}

The multiple-drive paradigm we propose here gives rise to the possibility of engineering emergent band models. These emergent bad structures could produce interesting and potentially useful dynamics of photonic systems as well as access new entangled states of photons and matter. Furthermore, they are not limited to the two-drive example presented in our manuscript.

The implementation of BHZ model in the Floquet space that was our focus here is just one example of a model that can be implemented by means of strong AC driving.
Many more exotic models can be built and analyzed by the same approach, by applying more drive frequencies (equivalent to going to higher-dimensional frequency spaces), or by pumping systems with more levels (including spatially extended ones). In particular, by applying three pump frequencies to a four level system, one may be able to construct the 3D extension for BHZ model \cite{FKM}. The topological invariant in this case is the second Chern number that represents  the magneto-electric response. Again, the ``electric field" is built into the Floquet model by construction.  It will be interesting to determine what observable the orbital ``magnetic'' response would corresponds to in the Floquet implementation.

Pumping spatially-extended systems by space-time dependent drives is another direction worth exploring. Mixing real-space with Floquet-space gives a simple interpretation to such classic effects as Thouless pump in terms of 2D quantum Hall effect (Appendix \ref{sec:TP}), and can also help discover new phenomena. E.g., 
by pumping 1D spatial lattice with two incommensurate frequencies one may be able to construct 3D topological insulators with $SU(2)$ Landau levels \cite{Congjun1, Congjun2}. By applying even more drives one may be able to access and study  more exotic yet states, such as eight-dimensional quantum Hall effect \cite{8dHall}.

\section{Acknowledgements}  

Authors would like to thank J. Sau, M. Sanchez, A. Yacobi, V. Manucharyan, M. Gullans, M. Lukin, F. Nathan, and Y. Oreg for useful discussions. IM acknowledges support from Department of Energy, Office of Basic Energy Science, Materials Science and Engineering Division. GR acknowledges support from NSF through DMR-1410435, as well as the Institute of Quantum Information and Matter, a NSF Frontier center funded in part by the Gordon and Betty Moore
Foundation, and the Packard Foundation. BH acknowledges support from STC Center for Integrated Quantum Materials, NSF grant DMR-1231319. This research was supported in part by the National Science Foundation under Grant No. NSF PHY-1125915. We are also grateful to the Aspen Center for Physics where part of the work was done. 

\appendix

\section{Weak drive}\label{App:weak}

When the drive is weak then the motion of the system in the Floquet space is highly restricted. This can be seen from
\beq
(E + \omega_1n + \omega_2 m)\phi^\alpha_{nm} = h^{\alpha\beta}_{pq} \phi^\beta_{n-p, m-q}.
\eeq
Due to large energy mismatches between neighboring sites, the ``particle" can hop on the lattice only along a 1D ``Manhattan" path that skirts the constant-energy line, e.g., $E + \omega_1n + \omega_2 m = 0$. The sites of the corresponding 1D model have onsite  energies given by 

$$u_{nm} = \omega_1n + \omega_2 m  = \sqrt{\omega_1^2 + \omega_2^2}(n \sin\alpha + m\cos\alpha) .$$
This onsite energy is proportional to the distance between a lattice point $(n,m)$ and the line cutting through the lattice at angle $\alpha$.  For $\tan\alpha$ irrational, this distance is a quasiperiodic quantity, and hence, the system represents a 1D lattice with quasiperiodic onsite potential. This can be compared to the well studied case of the Fibonacci quasicrystal \cite{Janot}. There, $\tan\alpha = \gamma$, the onsite energies are all the same, but the hoppings are different if they derive from the ``vertical" ($m$-direction) or the ``horizontal" ($n$-direction) hopping. Clearly, we can also introduce that by choosing $h_{01}\ne h_{01}$, however, the mapping to 1D only applies if $\omega_i >> |h|$, and thus it is not possible to exactly replicate the limit of pure bond ``disorder".
The spectrum of this problem is dense everywhere due to the symmetry $E\to E + \omega_1n + \omega_2 m$ accompanied by wave function translation.

\section{Commensurate drive frequencies -- detailed treatment.}\label{sec:cd}

When drives are incommensurate, the wavefunction $\phi_{\vec{n}}$ in Eq. (\ref{eq:psi1}) is defined on the whole 2D Floquet plane, since all frequencies $n\omega_1 + m\omega_2$ are unique. Suppose now that $\omega_1/\omega_2 = p/q$, with $p\lt q$ mutually prime integers. It is easy to see that if $p = 1$, then $n\omega_1$ exhausts all the possible frequencies. For $p\ne 1$, the number of independent frequencies is $p$ times that. Thus, in the $commensurate$ case the Floquet space is  ``compactified" into a strip of width $p$ along $\omega_1$ axis. The boundary conditions are periodic along $\omega_2$ direction, but with an ``offest" of $q$, introducing a shear (twist).

Let's see how this comes about in detail. As before, the Schr\"odinger equation is
\beq
i \partial_t \psi^\alpha(t) = H^{\alpha\beta}(t)\psi^\beta(t),\label{eq:SEc}
\eeq
with $H(t) = \cH( \omega_1 t, \omega_2 t)$, with $\cH(\theta_1, \theta_2)$  $2\pi$-periodic in $\theta_{1,2}$. Hence 

\bea\cH(  \omega_1 t,   \omega_2 t) = \sum_{\vec{m}} h_{\vec{m}}e^{ -i \vec{\omega}\cdot\vec{m} t}\nonumber\\
 = \sum_{``strip"} \tilde h_{\vec{m}}e^{ -i \vec\omega\cdot\vec m t}
\eea
where we combined all terms with the same frequencies, thus reducing summation to a strip of frequencies described above.

Similarly, the Floquet representation for the wave function is
\bea
\psi(t) = e^{-iEt }\phi(t) = \sum_{n,m} e^{-iEt - i \vec\omega\cdot\vec n t}\phi_{\vec n} \nonumber\\
= \sum_{``strip"} e^{-iEt - i \vec\omega\cdot\vec n t} \tilde\phi_{\vec n}\label{eq:psi}.
\eea
Substitution into the Schr\"odinger equation leads to the expected result: when action of Hamiltonian seems to take the system outside the strip, a shift by $\pm (-q,p)$ brings it back into the strip. This corresponds to the ``sheared" periodic boundary conditions in the strip. 

An alternative selection of unique frequencies is a tilted ribbon with the regular (non-sheared) periodic boundary conditions. The construction is the same as in the case of carbon nanotubes with helicity $(p,q)$. The Hamiltonian induces hopping on the ``nanotube"; the energy conserving dynamics corresponds to hopping along the nanotube circumference (perimeter $\sqrt{p^2 + q^2}$).

Note that if $p,q \gg 1$ then the compactification should not be noticeable, and one does not expect any difference between irrationally and such rationally related frequencies (e.g. if one constructs topological insulators in the Floquet space).

\section{Floquet eigenstates in quasiperiodic potentials}\label{App:floq}

When drives are quasiperiodic (irrational frequency ratio), Floquet theorem does
not directly apply, since one cannot define the evolution operator over the full period. Nonetheless, a quasiperiodic drive with irrational $\omega_2/\omega_1=\gamma$ can be approximated by a periodic one, $\omega_2/\omega_1= p/q$ ($p$ and $q$ are positive relatively prime integers). For instance, for $\gamma$ the Golden mean, $q$ and $p$ can be chosen as consecutive Fibonacci numbers. Then, the drive is periodic, with the period $T = pT_1 = qT_2$, and the Floquet theorem can be applied. Hence one can ask, whether in the limit of $p,q \to \infty$, the Floquet eigenstates (FE) converge to unique states. These could be naturally defined as the Floquet eigenstates for the case of incommensurate drives. 

For a periodic drive, FE have the form $\Psi(t) = e^{-iEt} \Phi(t)$,
with $\Phi(t)$ periodic. The FE are the eigenstates of the evolution operator over the period, 

$$U(T) = {\cal T} \exp-i\int_0^T H(t')dt'.$$

For a TLS, this is a 2x2 matrix, that can be easily computed and diagonalized to find FE.

\begin{figure}
\includegraphics[width=1\columnwidth]{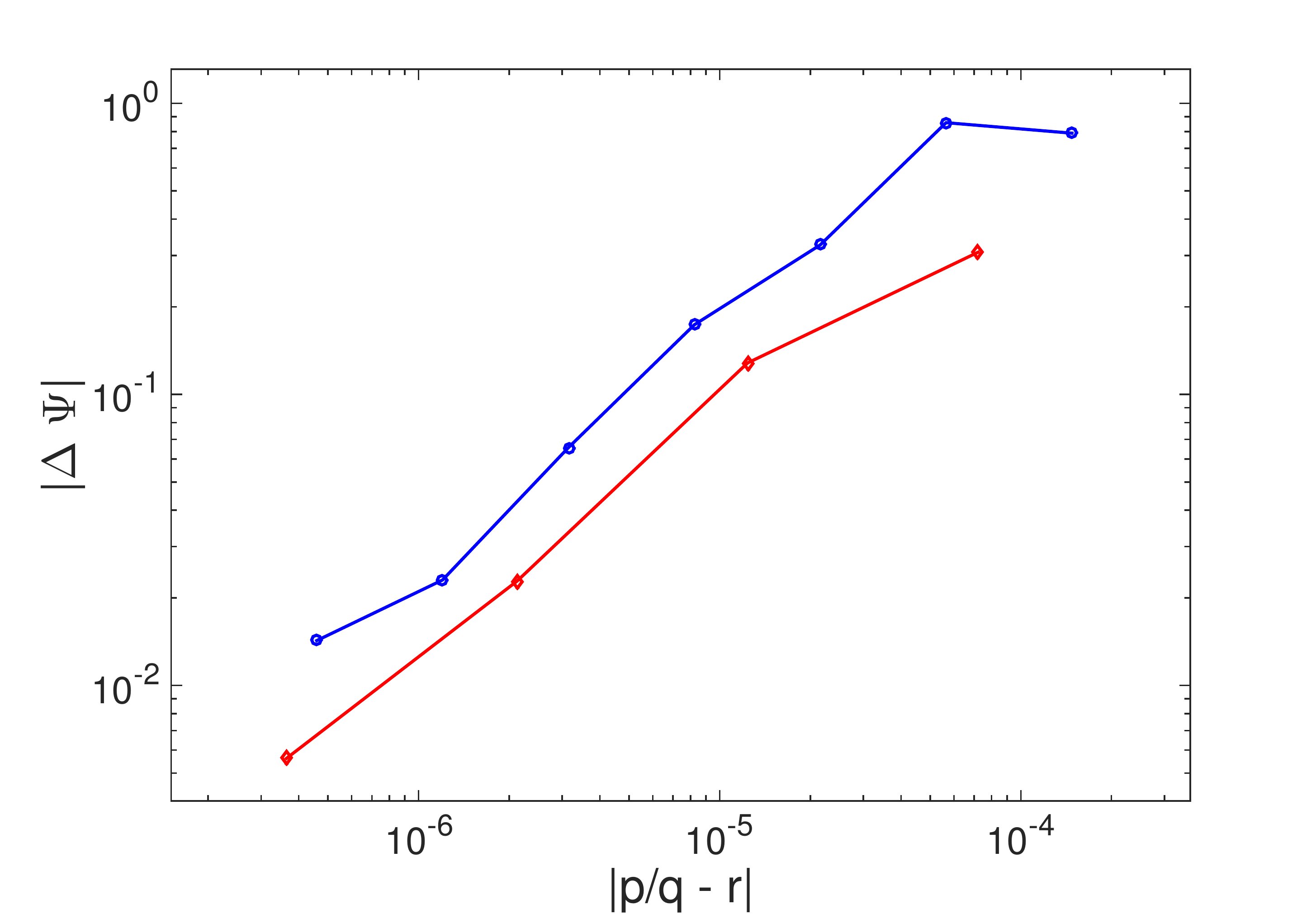}
\caption{Convergence of Floquet eigenfunctions for $\omega_1/\omega_2 = p_i/q_i \to r$, where $r$ is irrational [the Golden mean, $\gamma = 1.6180...$ (blue line) and $\sqrt{2}$ (orange line) ]. $|\Delta \Psi|$ is calculated relative to the best rational approximation used ($2584/1597$ for $\gamma$ and $3363/2378$ for $\sqrt{2} $); it is approximately linear in $ |p/q -r|$. Parameters of the Hamiltonian are described in the text.}\label{fig:FES}
\end{figure}

\subsubsection{Observations}

To test convergence of Floquet eigenstates, we studied Hamiltonian Eq. (\ref{eq:2fs}) with parameters $\eta = 1$ and $m = 1.5$.  We considered  $\omega_2/\omega_1 = \lim_{i\to \infty}p_i/q_i = r$, with irrational $r$ chosen as $\gamma$  (with $\omega_1 = 0.3$) and $\sqrt{2}$ (with  $\omega_1 = 0.5$). We used continued fraction representations of
$\gamma = \frac{1}{1+\frac{1}{1 + \frac{1}{1+ ...}}}$ and $\sqrt{2} = 1 + \frac{1}{2+\frac{1}{2+\frac{1}{2+ ...}}}$
terminated at a finite level to obtain the sequences of (the best \cite{Hurwitz}) rational approximations. The largest-denominator approximations that we considered were $2584/1597$ for $\gamma$ and  $3363/2378$ for $\sqrt{2}$.  
In Figure \ref{fig:FES} the deviation, $|\Delta \Psi|_i = |\Psi_{p_i/q_i}-\Psi_{p_{\max}/q_{\max}}|$ is plotted as a function of $|p_i/q_i - r|$ for the two cases [the reference values are $\Psi_{2584/1597}  \approx (0.094 + 0.496i,   0.863)^\dag$ and $\Psi_{3363/2378} \approx (-0.230 - 0.446i,
  0.865)^\dag$]. Both show approximately linear convergence, indicating the existence of the  Floquet eigenstates for incommensurate drive frequencies.  
We also find, as expected, that in the limit of small frequencies ($\omega_i \ll \eta$) the FE are nearly the eigenstates of the instantaneous Hamiltonian. 

Note, that we only tested the FE convergence in the gapped phase. While we have not tested it explicitly, we expect that in the gapless points ($|m| = 0,2$), the limit may not exist, since small changes in frequencies would lead to topologically distinct trajectories in the Floquet zone.

\section{Quantum pump analogy}\label{sec:QP}

It is tempting to rewrite the Berry curvature of Eq. (\ref{eq:BC}) as
\be
 \Omega_{\vec{q}}=\hat{z} i\frac{1}{2\epsilon_{\vec{q}}^2}\mbox{tr}\l(P_{\vec{q}}\l[\der{\H_{\vec{q}}}{q_1},\der{H_{\vec{q}}}{q_2}\rr]\rr),
 \ee
(easy to see since $\mbox{tr}\l(P_{\vec{q}}\H_{\vec{q}}\der{P_{\vec{q}}}{q_i}\rr)=0$).
This leads to a formula for the pump power
\be
\frac{1}{2}\der{(E_1-E_2)}{t}= \frac{\omega_1\omega_2}{2\epsilon_{\vec{q}}^2}\mbox{tr}\l(P_{\vec{q}}\l[\der{\H_{\vec{q}}}{q_1},\der{H_{\vec{q}}}{q_2}\rr]\rr).\label{QP}
 \ee
Eq. (\ref{QP}) bears close similarity to the quantum pumping formula \cite{Brouwer}, except that instead of the $S$-matrix, the derivatives in the commutator are of a Hamiltonian. Indeed the photons that enter and leave the driven system do not have a conservation law (since they have different frequencies), and therefore they can not be assigned an $S$-matrix. This equation may suggest that a generalization of the quantum pumping formula for conserved but non-quantized quantities, such as energy, may exist.
 
\section{Floquet vs Instantaneous eigenstate initialization}\label{sec:FvsI}

\begin{figure*}
a.\includegraphics[width=.85\columnwidth]{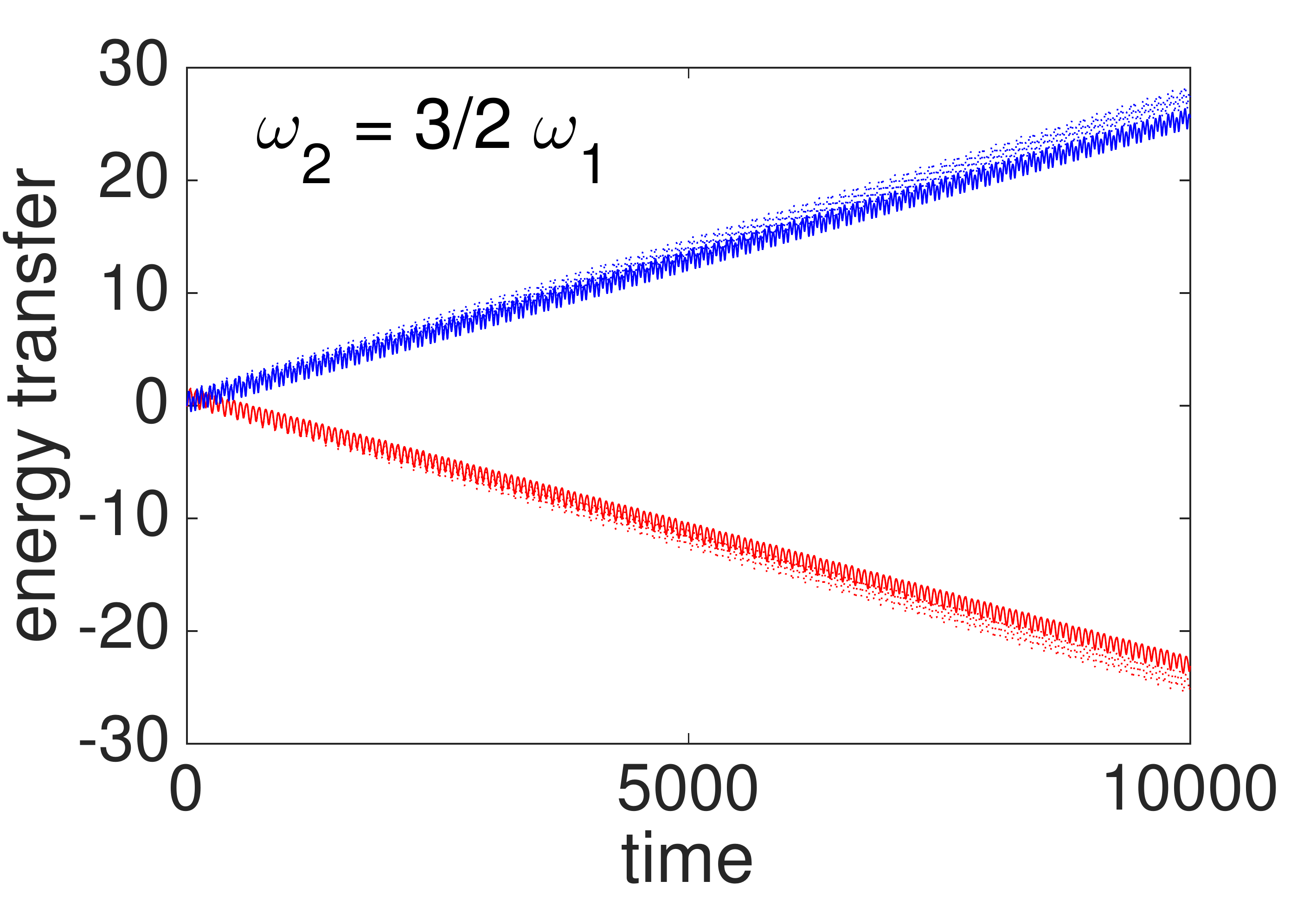}
b.\includegraphics[width=.85\columnwidth]{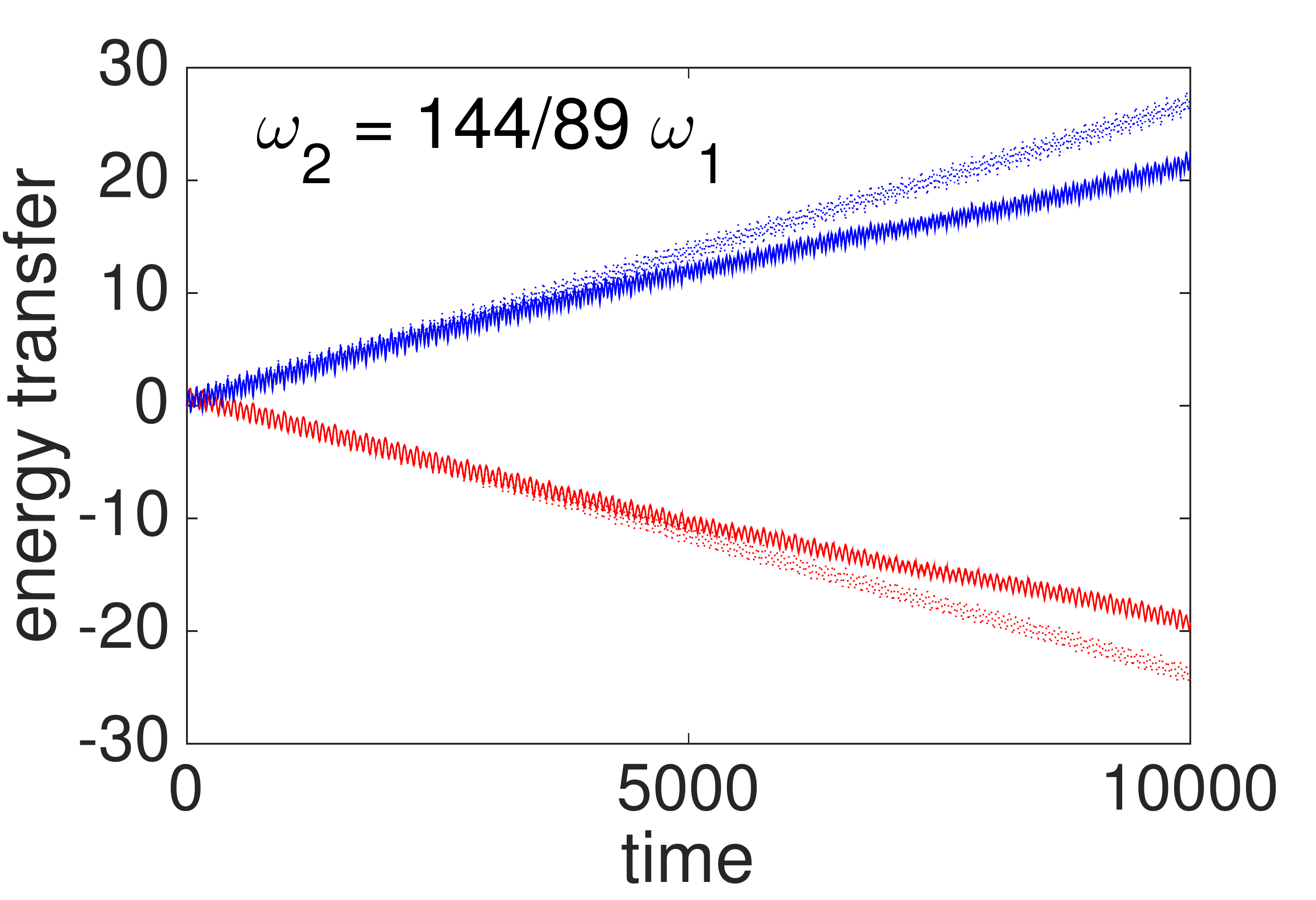}
c.\includegraphics[width=.85\columnwidth]{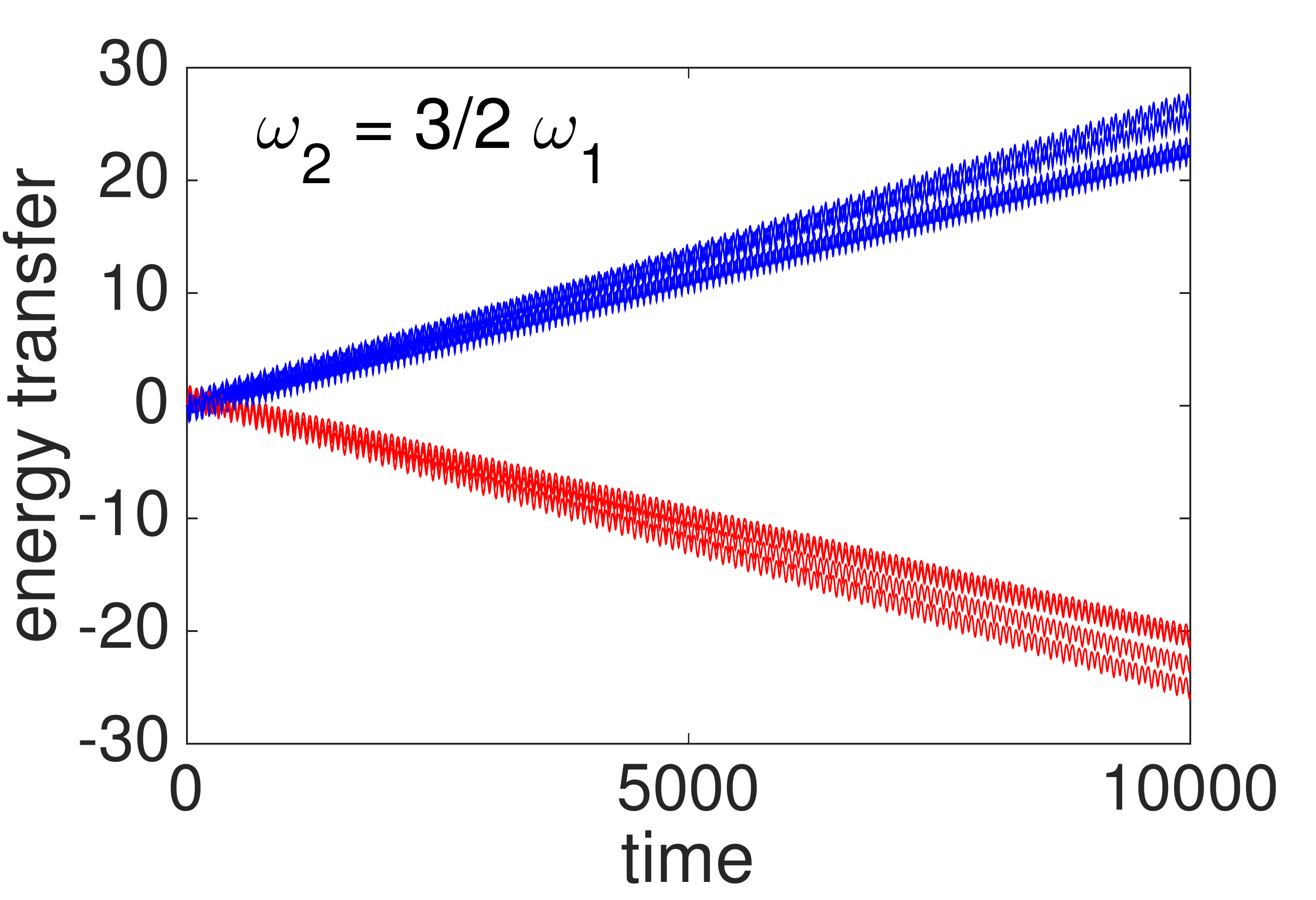}
d.\includegraphics[width=.85\columnwidth]{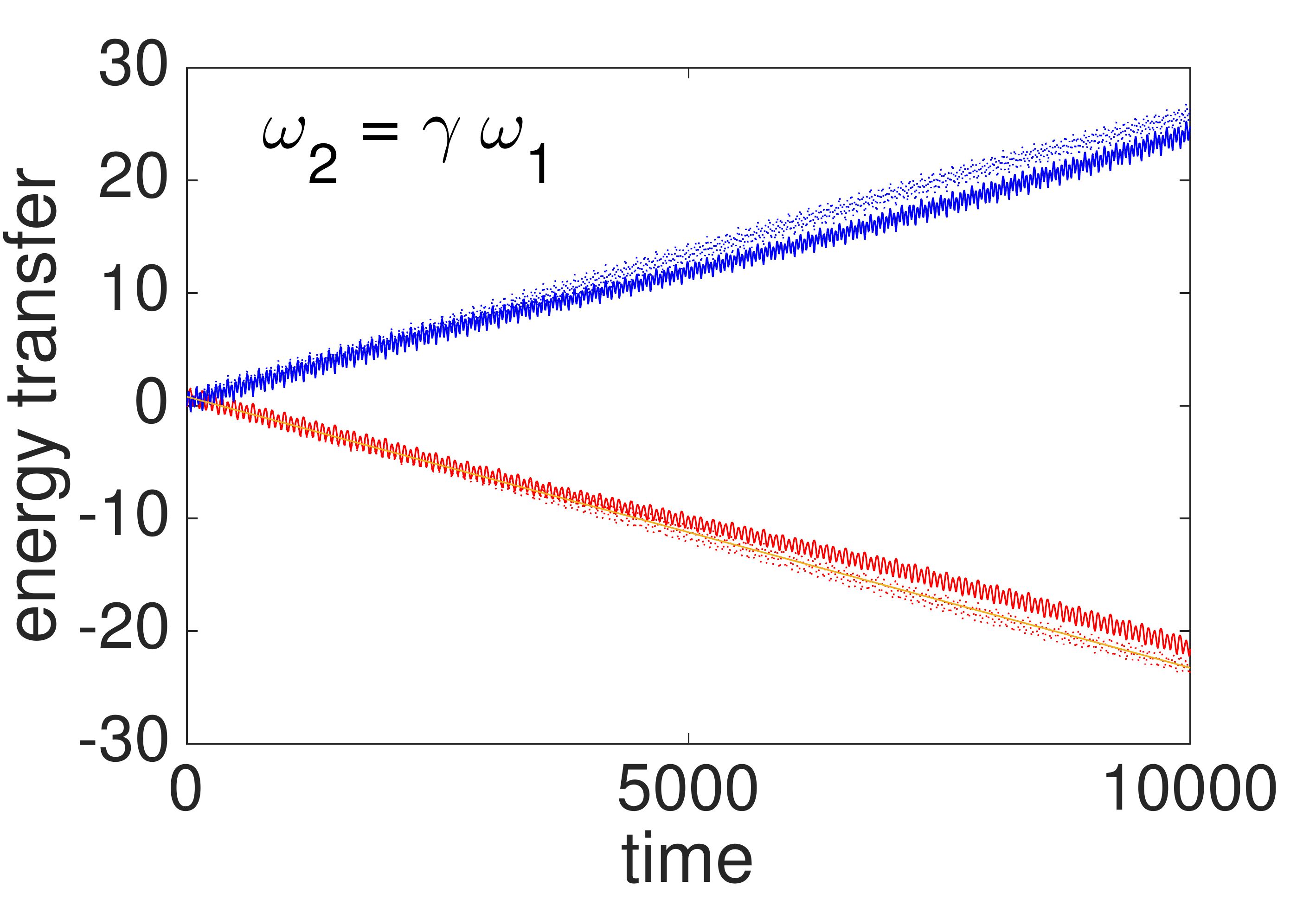}
\caption{Comparison of evolution starting from instantaneous (solid lines) and Floquet (dotted lines) eigenstates. Rational frequencies in a-c are approximations of the Golden mean $\gamma$. Hamiltonian scale is $\eta = 1$ and $\omega_1 = 0.1$, with $\omega_2$ stated on plots. Panel c illustrates how the energy transfer for rational drive depends on the the relative phase, $\phi_2 - \phi_1 = 0, 1.5, 3, 4.5$, starting from the respective instantaneous eigenstates. The variation is due to sampling of different closed lines in the phase Brillouin zone. Panel d illustrates how initializing into a Floquet eigenstate (computed for $\omega_2 = 144/89\omega_1$), enhances pumping at long times for irrational drive.
}\label{fig:CICh1}
\end{figure*}

In Figure \ref{fig:CICh1} we consider two rational approximations to the Golden mean: $\omega_2  = 3/2\omega_1$ and $\omega_2  = 144/89\omega_1$, for $\eta = 1$ and $\omega_1 = 0.1$. 
First of all, even in the case of  small commensuration, $\omega_2/\omega_1 = 2/3$ there is significant pumping. The rate of pumping is  similar to the one for a nearby incommensurate ratio.  As mentioned in Section \ref{sec:com}, for the commensurate drives, the Floquet zone is not fully sampled, and hence the results can depend on the relative phase between the drives. Indeed, varying the initial phase shift, does lead to different pumping rates, Fig. \ref{fig:CICh1}c. 

In Fig. \ref{fig:CICh1}a,b,d we have compared the initializations into the instantaneous eigenstates (solid lines) and the Floquet eigenstates (dotted lines). In all cases we considered, we found that initialization into the Floquet eigenstates leads to faster pumping.
For incommensurate drive frequencies, Fig. \ref{fig:CICh1}d, the initialization was into the Floquet eigenstate of $\omega_2  = 144/89\omega_1$ (approximation to the incommensurate FE discussed in Appendix \ref{App:floq}). That lead to faster and more linear pumping, with the value near the theoretical upper limit Eq. (\ref{eq:P12}). The reason for faster pumping can be traced to higher average fidelity in the FE.

\begin{figure*}
a.\includegraphics[width=.85\columnwidth]{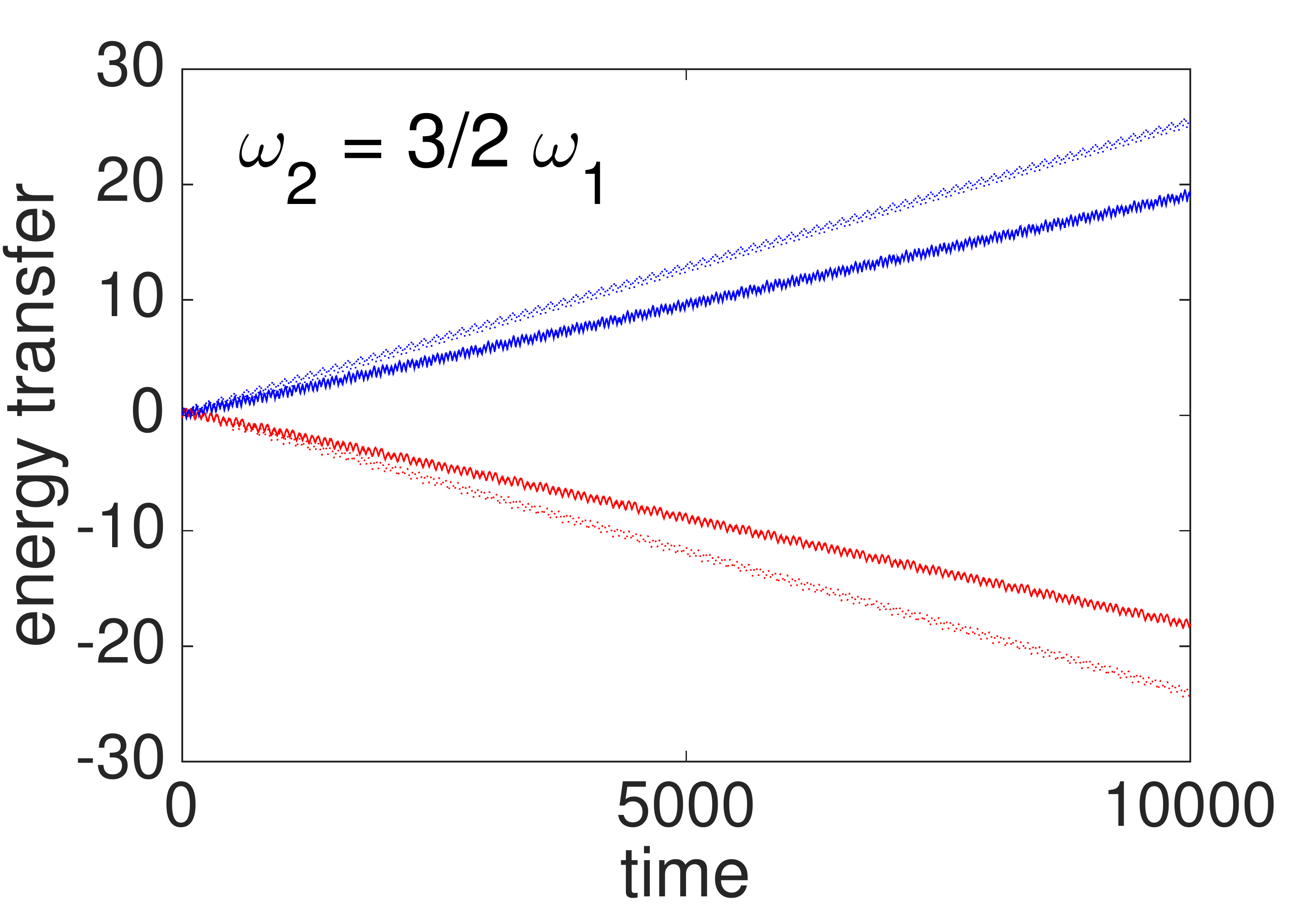}
b.\includegraphics[width=.85\columnwidth]{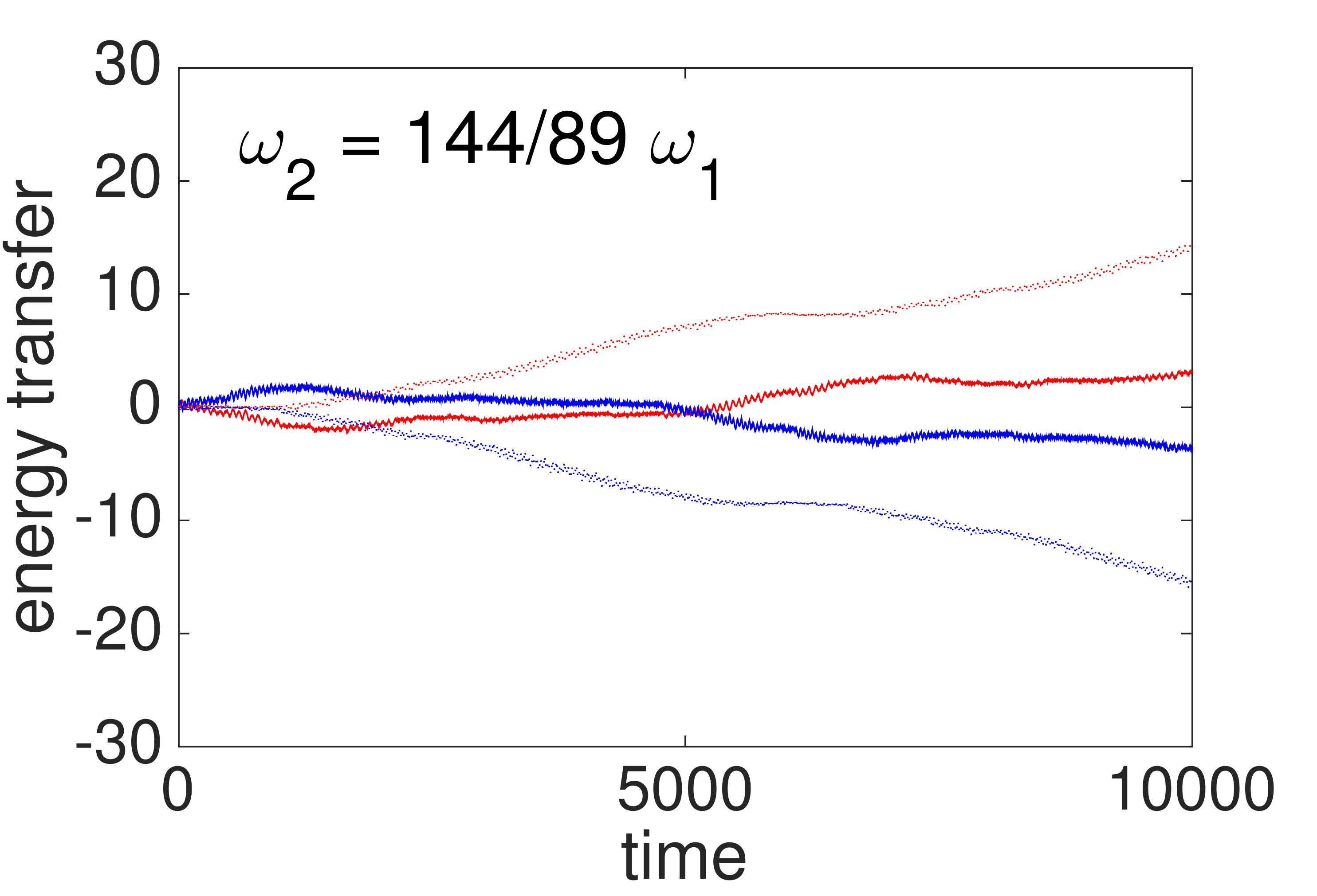}
c.\includegraphics[width=.85\columnwidth]{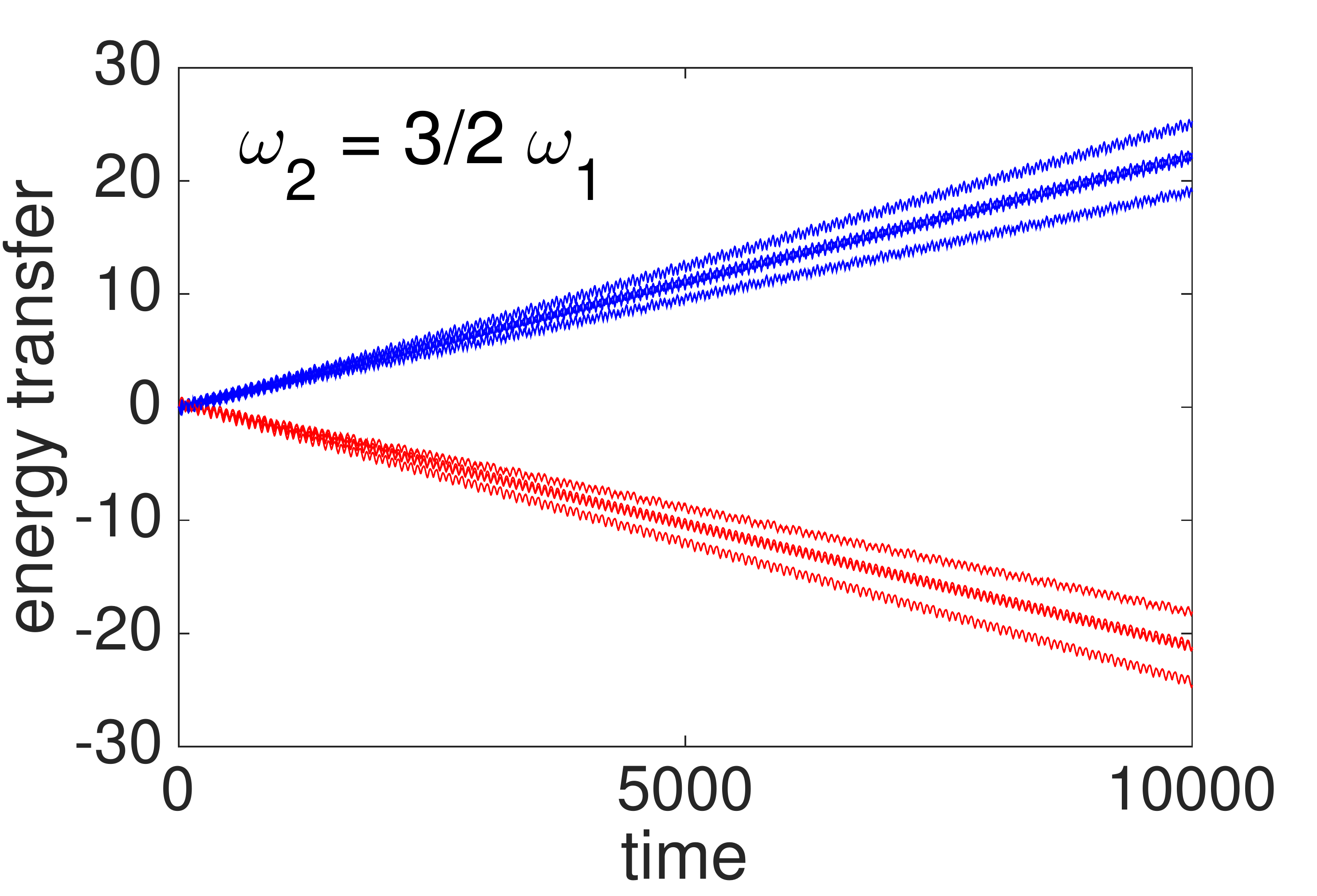}
d.\includegraphics[width=.85\columnwidth]{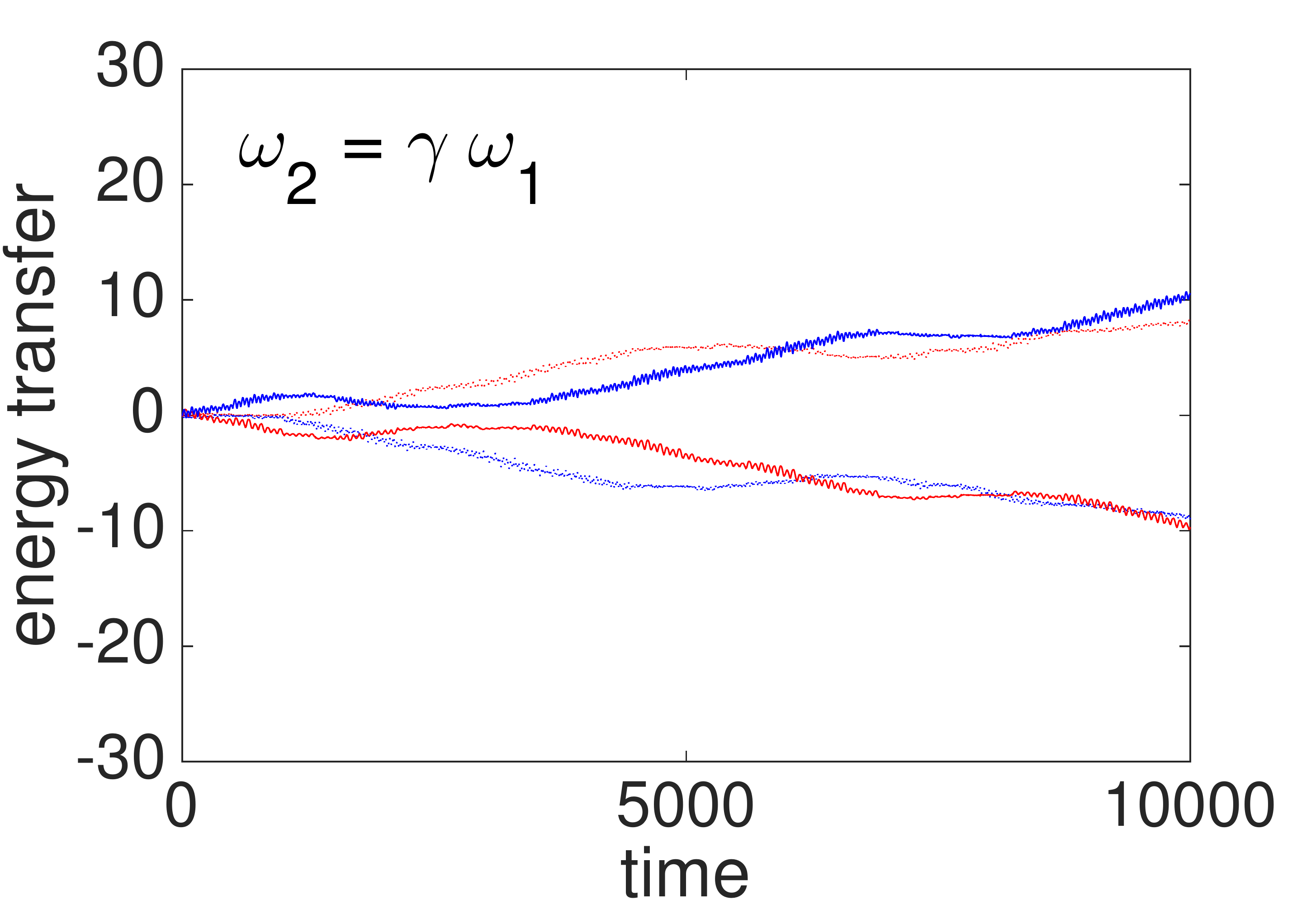}
\caption{Same as Figure \ref{fig:CICh1}, but for $\eta = 0.5$.
}\label{fig:CIChp5}
\end{figure*}

\begin{figure*}
a.\includegraphics[width=.85\columnwidth]{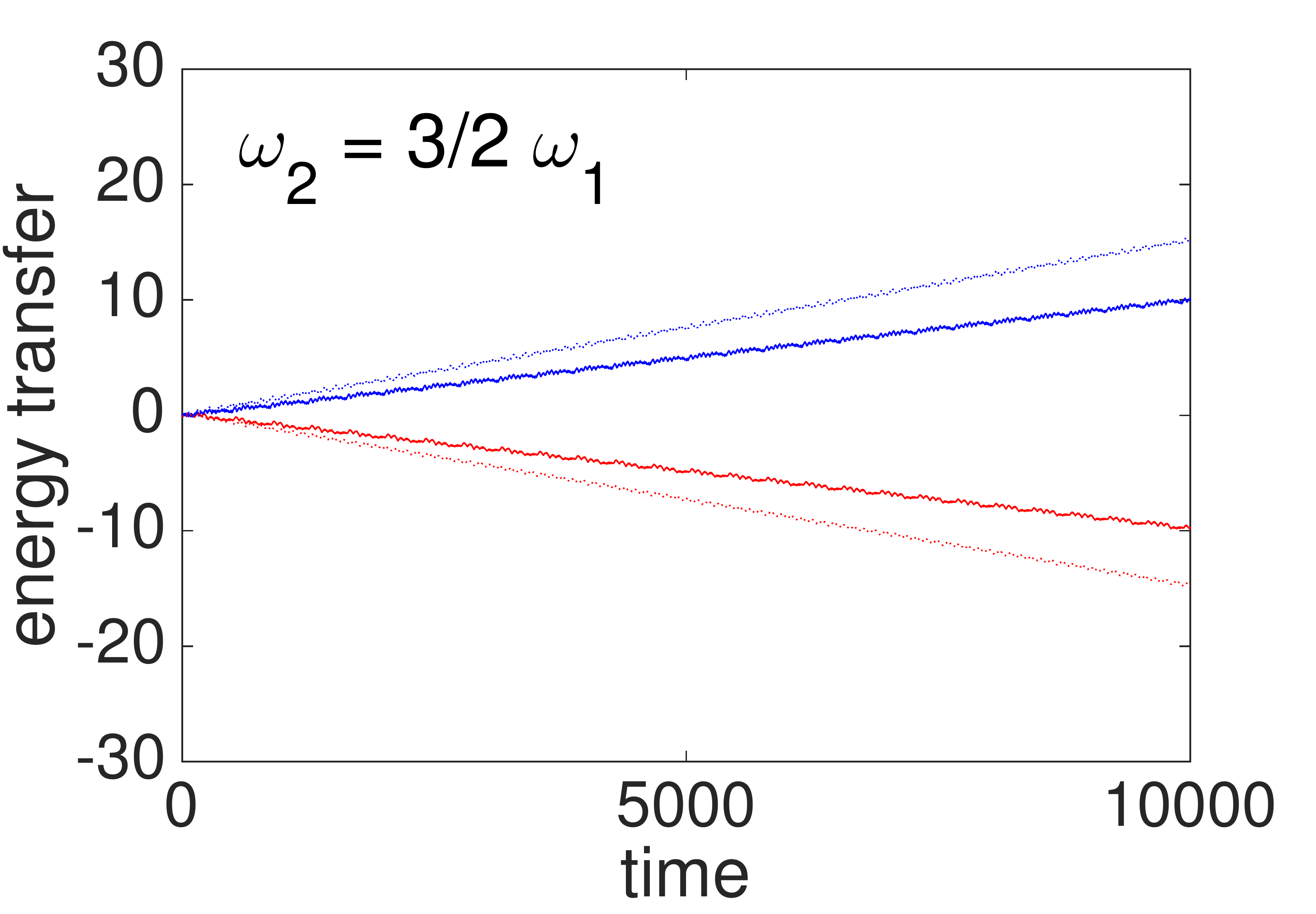}
b.\includegraphics[width=.85\columnwidth]{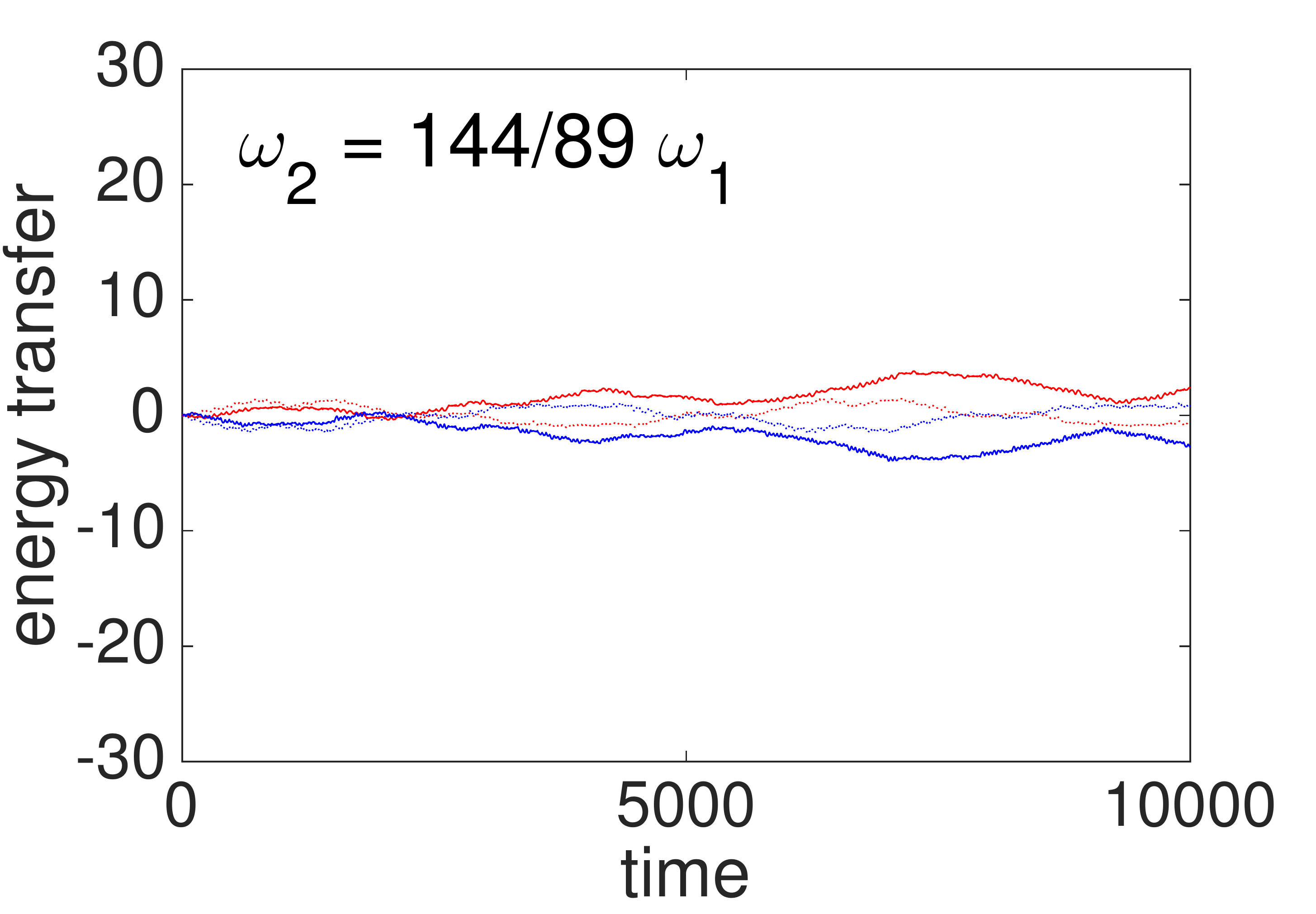}
c.\includegraphics[width=.85\columnwidth]{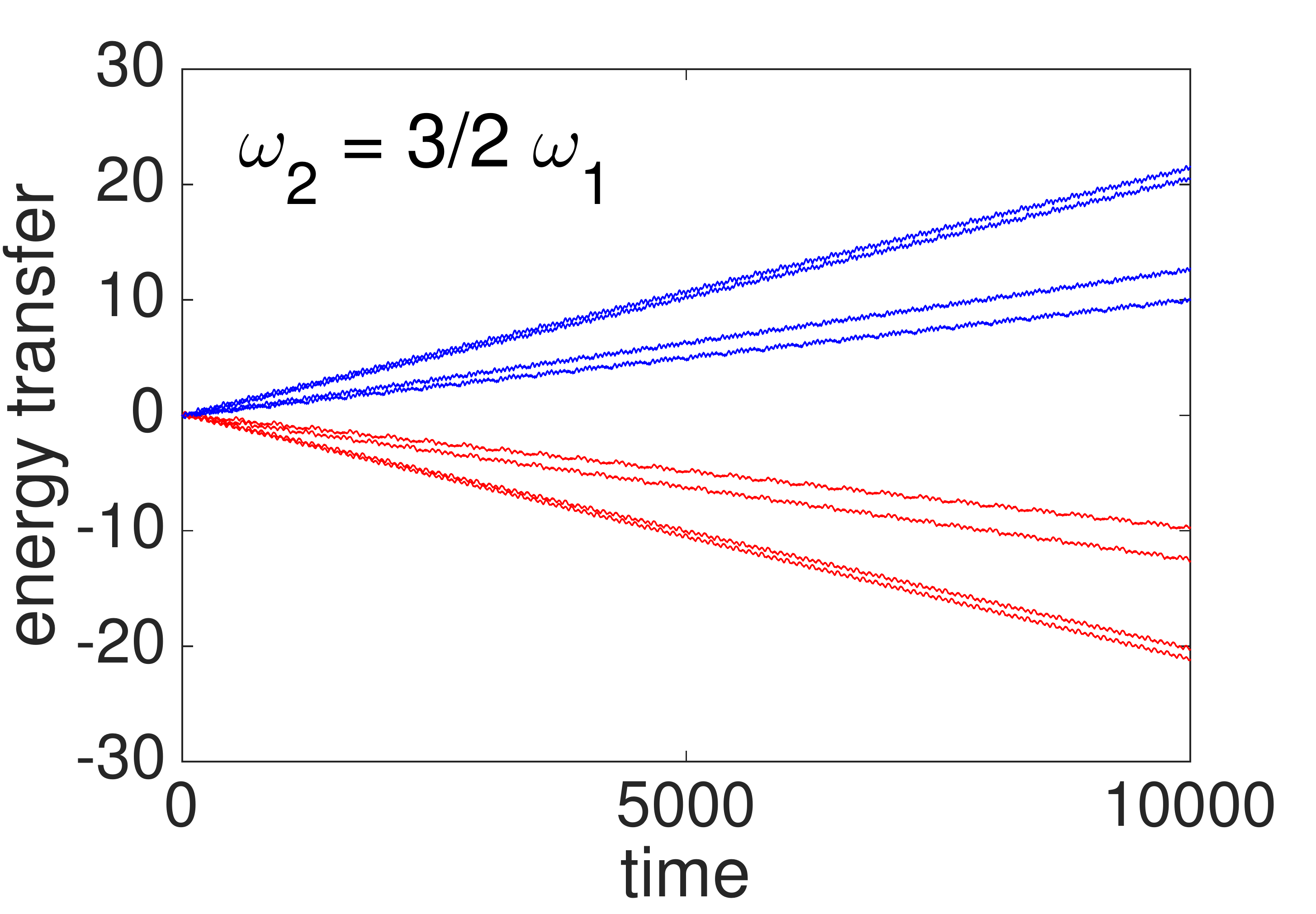}
d.\includegraphics[width=.85\columnwidth]{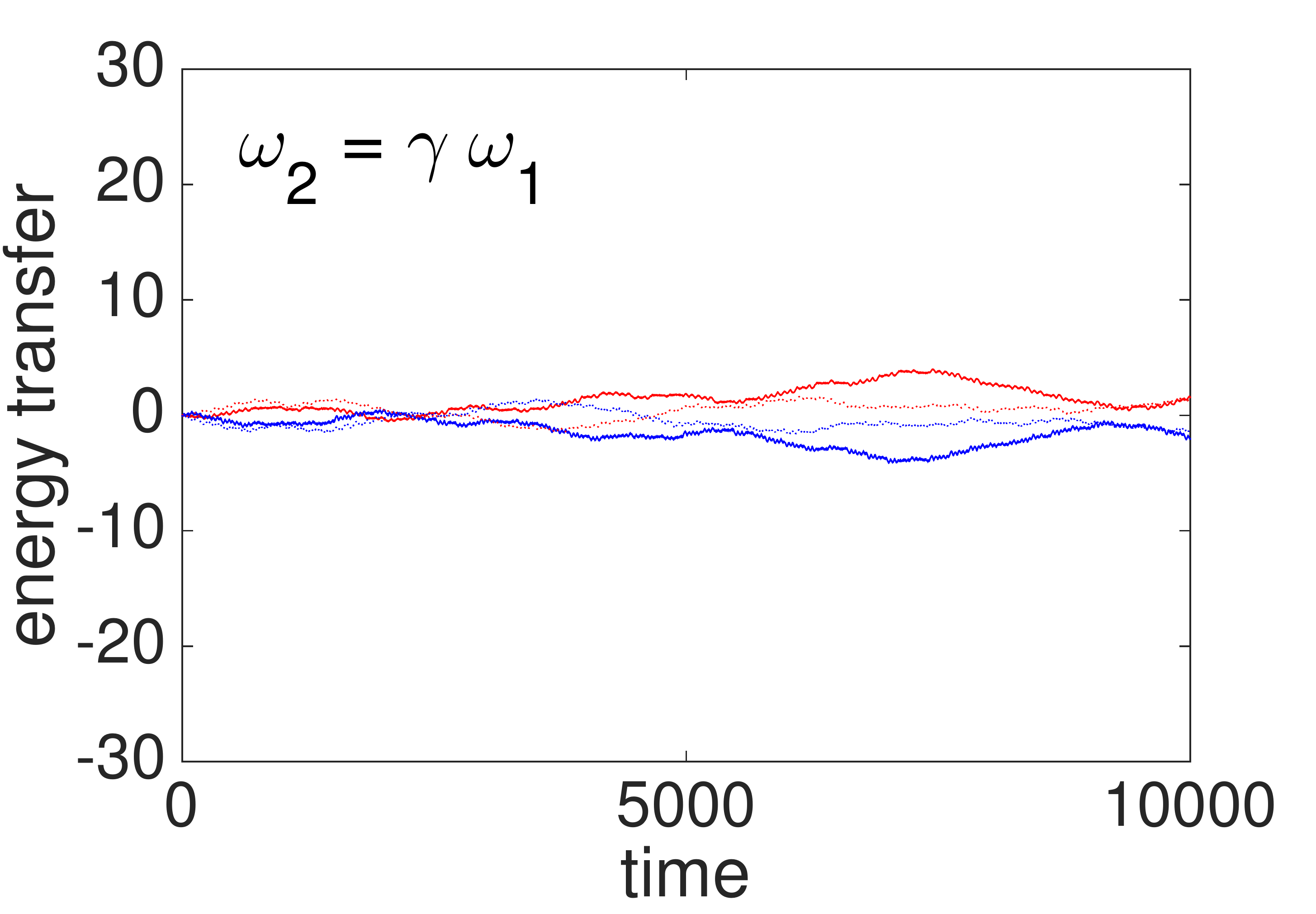}
\caption{Same as Figure \ref{fig:CICh1}, but for $\eta = 0.2$.
}\label{fig:CIChp2}
\end{figure*}

In Figures \ref{fig:CIChp5} and \ref{fig:CIChp2} we consider the energy pumping, for the same parameters as in Figure \ref{fig:CICh1}, but for reduced Hamiltonian scales, $\eta = 0.5$ and $\eta = 0.2$, respectively. Consistently with the expectation, the energy pumping is suppressed, particularly strongly for larger denominator (and irrational) fractions $\omega_1/\omega_2$. Interestingly, for small denominators, e.g. $2/3$, the pumping persists even for relatively small $\eta$.  

\section{Additional numerical results }\label{sec:moreNums}

In addition to the numerical results we presented in the text, we explored the intermediate and weak drive regimes of the double-drive BHZ model. We present these numerical results here for completeness. Unless stated otherwise, $\omega_1 = 0.1$ and $\omega_2 = \gamma \omega_1$, where $\gamma$ is the Golden mean.

\subsection{Intermediate drives, $\eta=0.5,1$}

\begin{figure}
a.\includegraphics[width=.90\columnwidth]{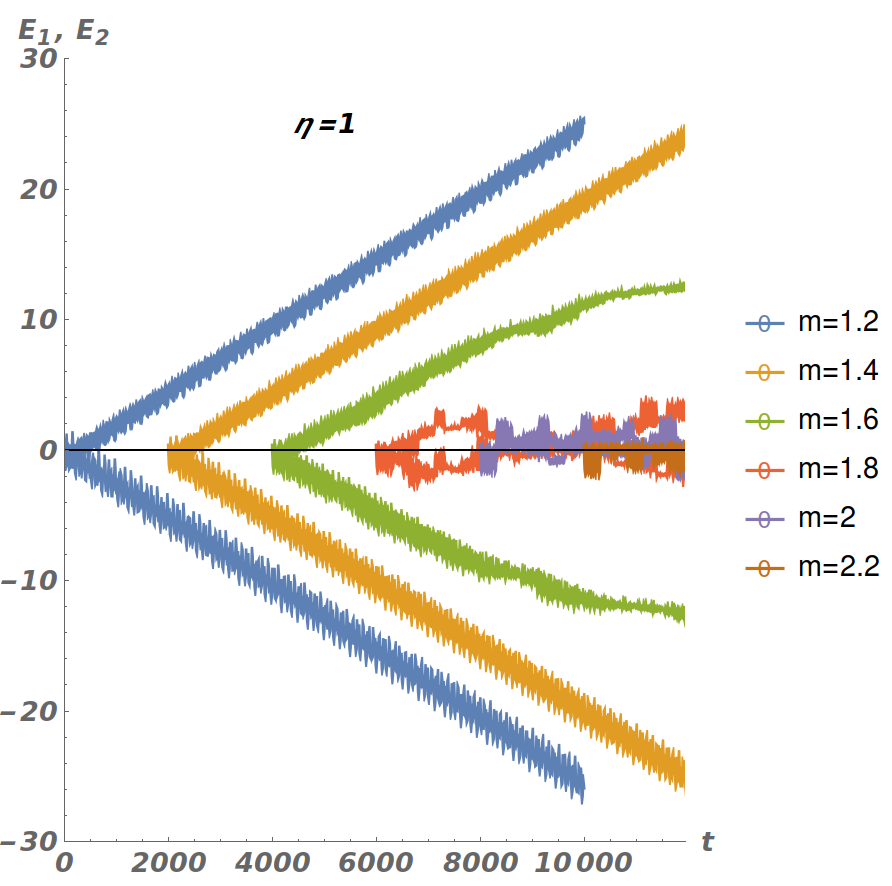}
b.\includegraphics[width=.90\columnwidth]{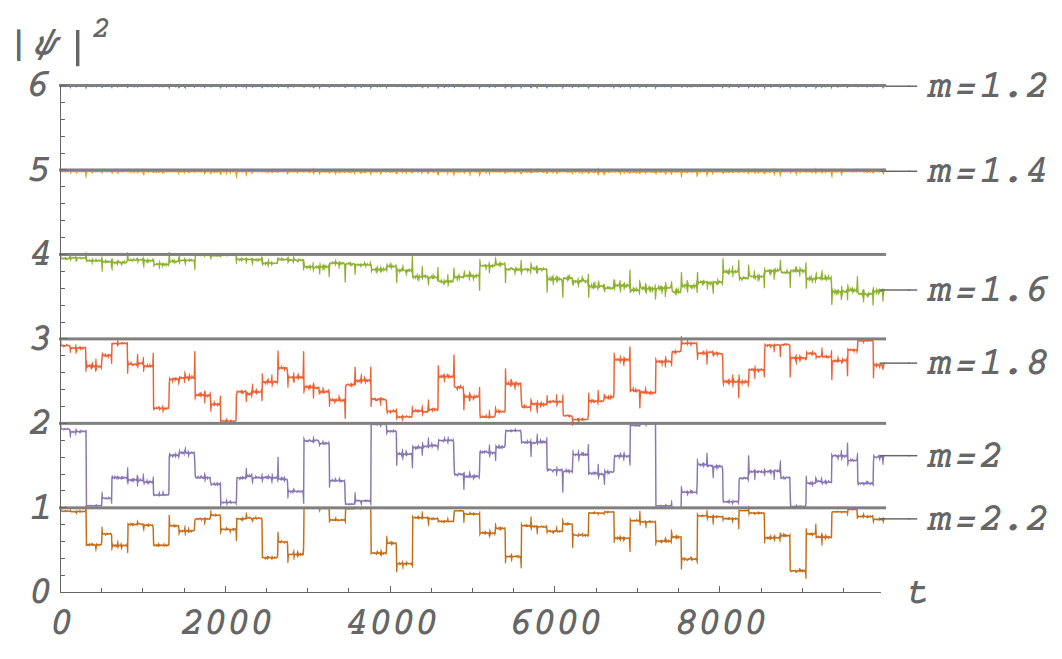}
c.\includegraphics[width=.90\columnwidth]{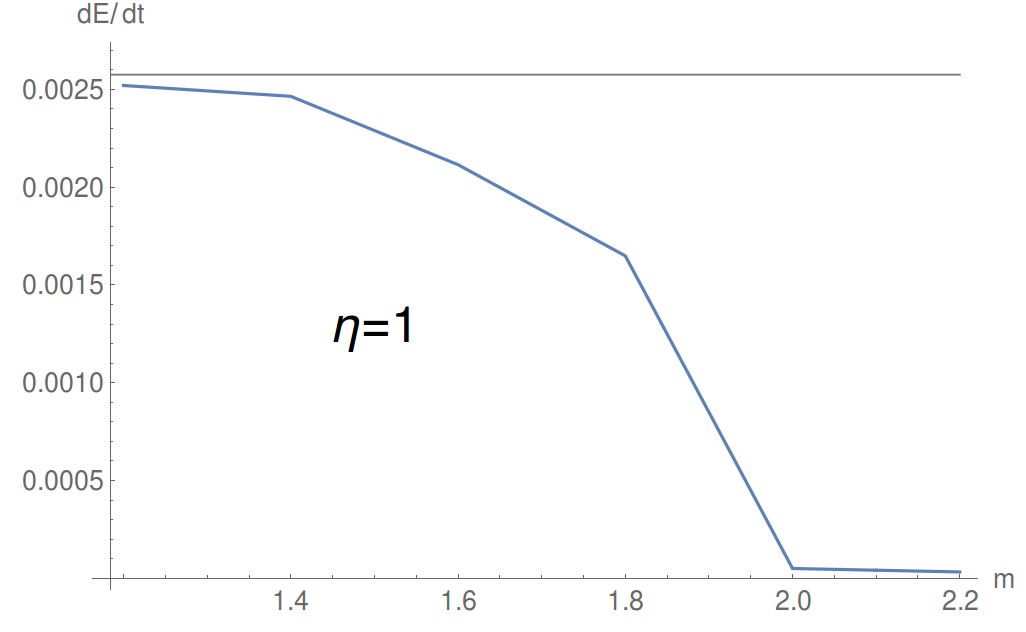}
\caption{Work (a)  and fidelity (b) in the intermediate drive regime for $\eta=1$. We initialized the system with the instantaneous eigenstate of $\H(0)$,  where the initial phases for these plots are $\phi_1=\pi/10,\,\phi_2=0$. 
(c) Power pumped averaged for $t<2000$  as a function of gap parameter $m$.
\label{int-flow1}}
\end{figure}

\begin{figure}
a.\includegraphics[width=.90\columnwidth]{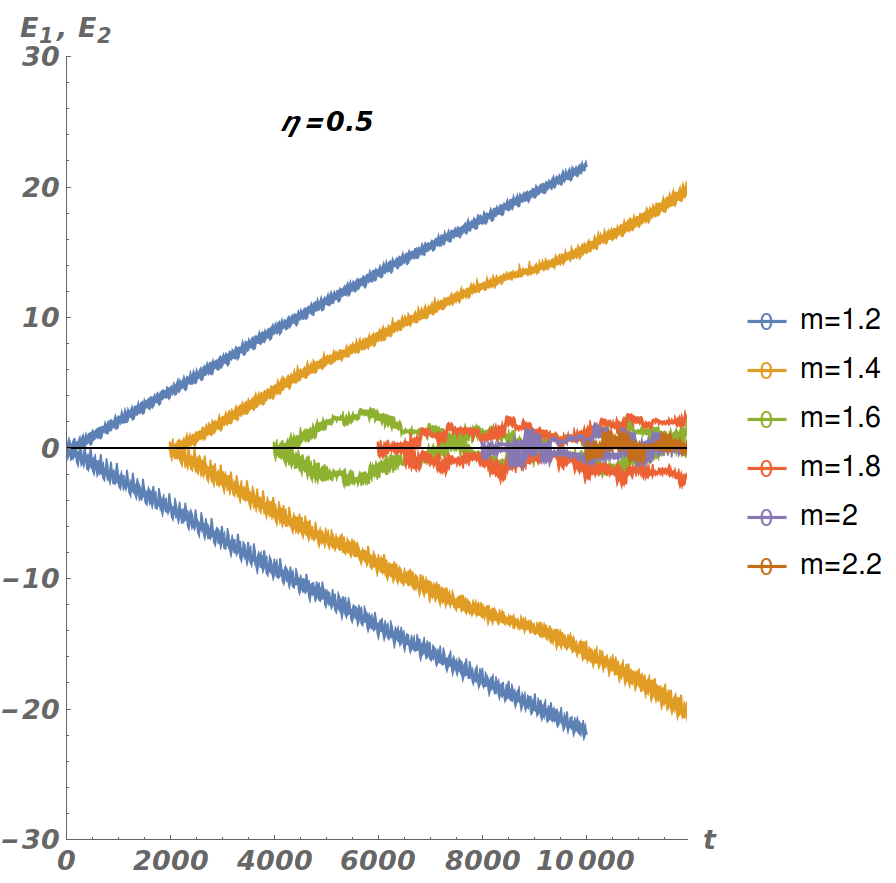}
b.\includegraphics[width=.90\columnwidth]{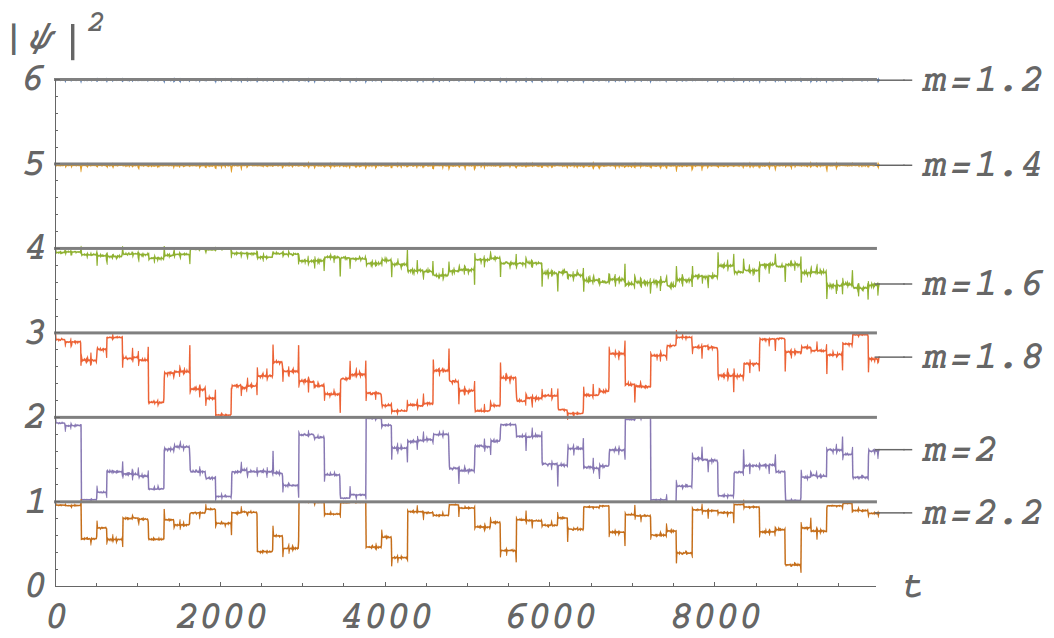}
c.\includegraphics[width=.90\columnwidth]{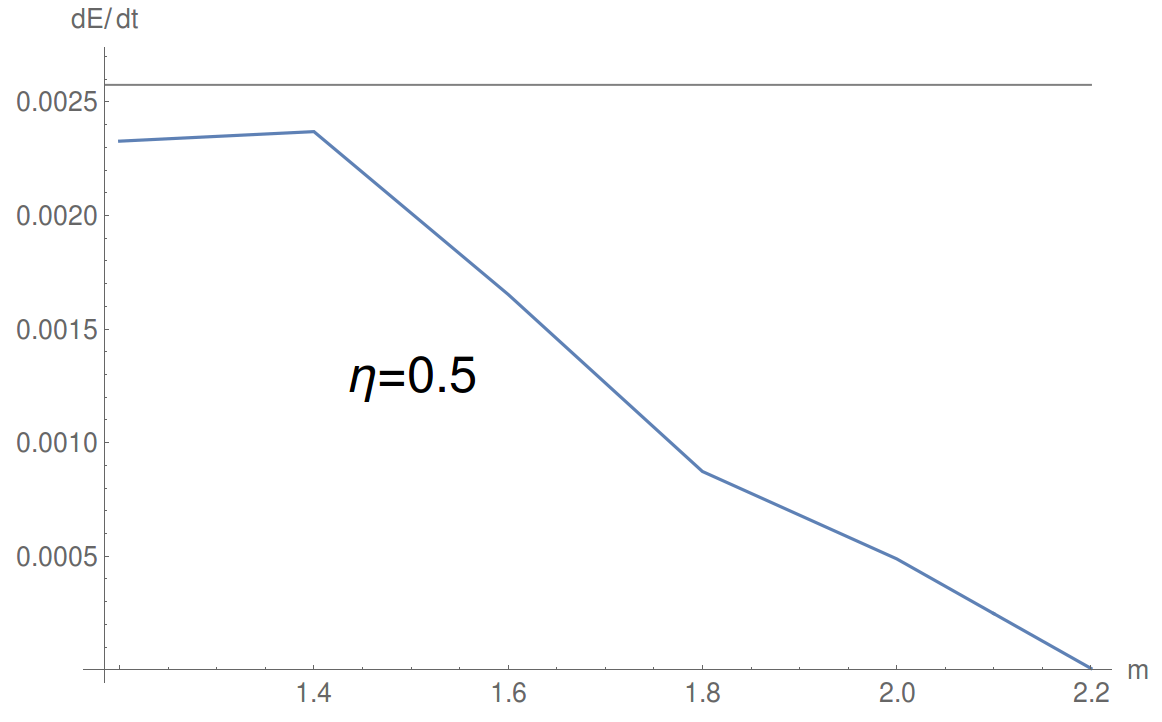}
\caption{Work (a) and fidelity (b) in the intermediate drive regime for $\eta=0.5$. We initialized the system with the instantaneous eigenstate of $\H(0)$,  where the initial phases for these plots are $\phi_1=\pi/10,\,\phi_2=0$. (c) Power pumped averaged for $t<2000$  as a function of gap parameter $m$.\label{int-flow2}}
\end{figure}

For $\eta$ values that exceed $\omega_{1,2}$, but not by much, we still see a strong pumping effect deep in the topological regime. The effect subsides, however, well before the phase boundary between the topological and trivial parameter regimes (Fig. \ref{int-flow1}a and \ref{int-flow2}a). As can be seen from Fig. \ref{int-flow1}b and \ref{int-flow2}b, this is a result of the fidelity being lost after a finite time in parameter ranges close to or beyond the topological transition into the trivial range. This is associated with the system exploring parts of the Floquet zone where the band gap of the underlying BHZ model is comparable to the drive frequencies.

Figs. \ref{int-flow1}c  \ref{int-flow2}c show the power absorbed by the drives averaged over the first 2000 time units of the evolution as a function of the gap parameter $m$ for the $\eta=0.5,\,1$. The transition between the pumping regime and the trivial regime is quite abrupt.

\subsection{Weak drives, $\eta\leq \omega_{1,2}$}

\begin{figure}
a.\includegraphics[width=.95\columnwidth]{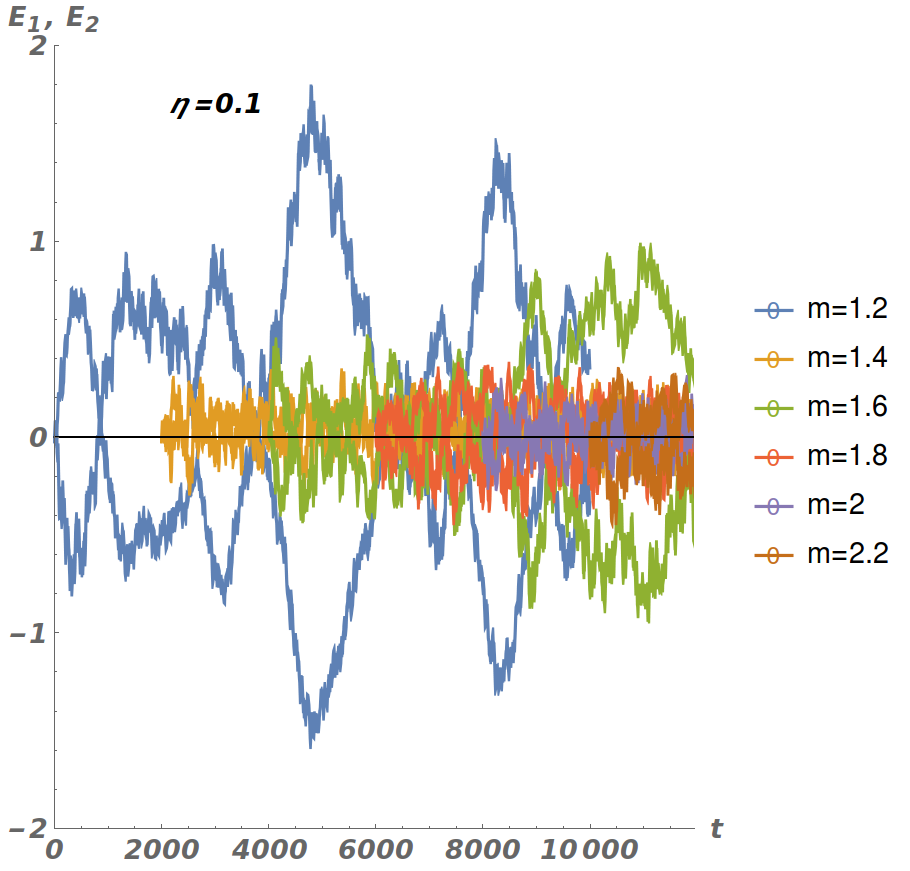}
b.\includegraphics[width=.95\columnwidth]{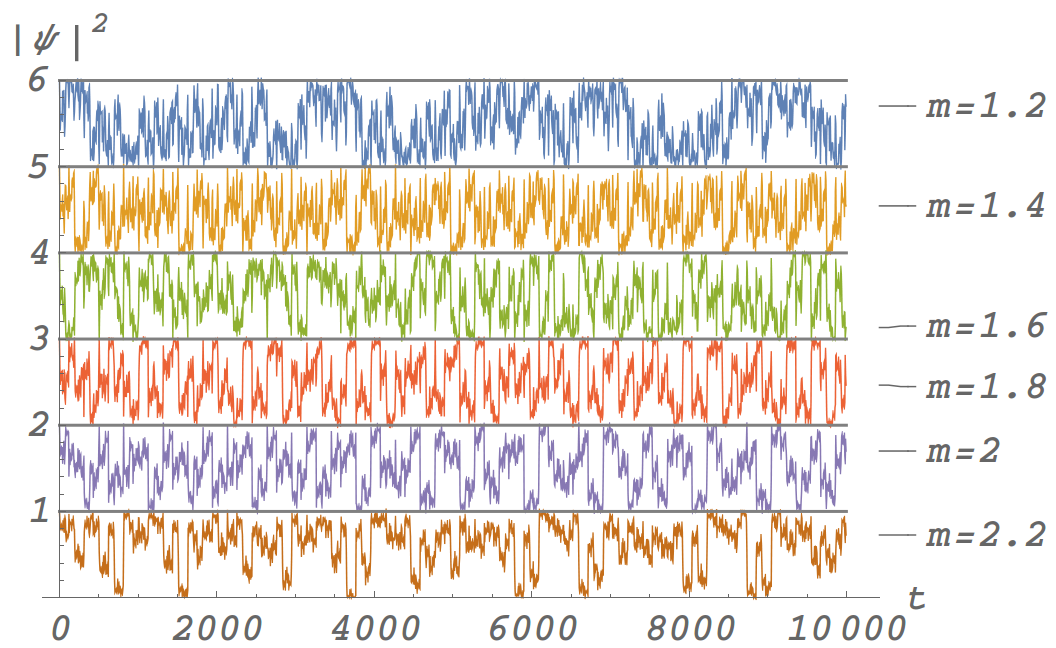}
c.\includegraphics[width=.95\columnwidth]{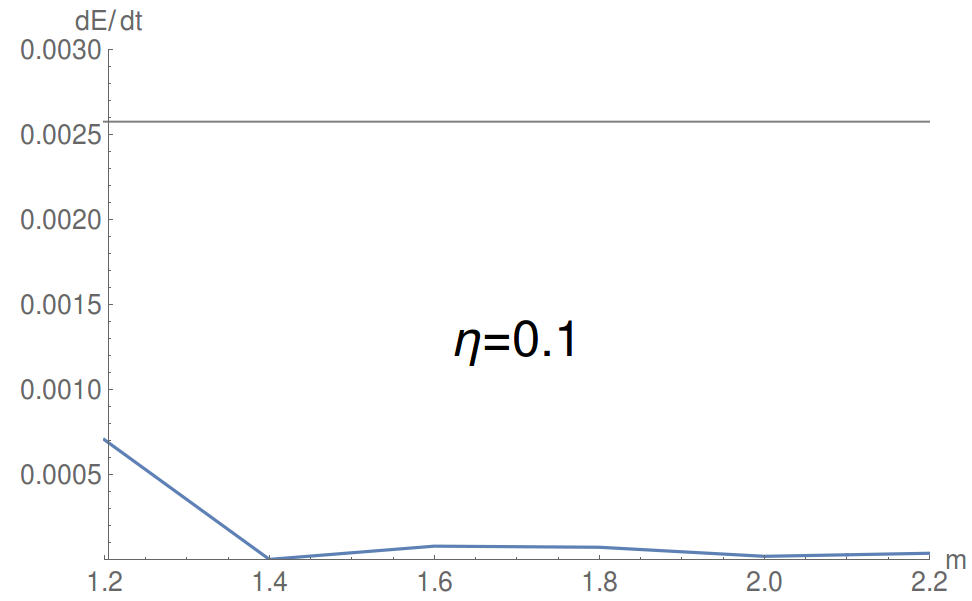}
\caption{Work and fidelity in the weak drive regime. (a) Energy transfers for $\eta=0.1$. (b) Fidelities of initial state, an instantenous eigenstate of $\H(0)$, and the instantaneous eigenstates of $\H(t)$.  The initial phases for these plots are $\phi_1=\pi/10,\,\phi_2=0$. (c) Power pumped averaged for $t<2000$  as a function of gap parameter $m$.\label{weak-flow}}
\end{figure}

To explore the weak-drive regime we considered the $\eta=0.1$. As Figures \ref{weak-flow} suggest, initializing the system with the instantaneous eigenstate of the Hamiltonian $\H(0)$ results in negligible pumping. Initialization into a Floquet eigenstate of a periodic approximation does not qualitatively change the result, as can be seen from Fig. \ref{fig:CIChp2}d.

\section{Thouless pump, reloaded}\label{sec:TP}

In this section we will provide the reinterpretation of the Thouless  pump \cite{TPump} using the Floquet lattice construction. Thouless pump is a classic example of a driven system that shows quantized transport behavior.
Consider a one-dimensional tight binding lattice, $x_i = i$,  with spatially- and time-dependent onsite potential, $V(x,t) = A \cos(\omega_0 t - k x_i)$.  The potential is time-periodic, and thus we can take advantage of the Floquet transformation described in Section \ref{sec:2w}. As a result we obtain a two-dimensional lattice -- one spatial dimension that no longer has spatially varying potential, and one frequency dimension, with the superimposed linear potential corresponding to an applied electric field of strength $\omega_0$. The hopping in the positive (negative) frequency direction is $A e^{\pm i k x_i}$. The phase factor corresponds to an effective linear in $x$ vector potential pointing in the $y$ (frequency) direction. This is equivalent to a uniform magnetic field of strength $k$ piercing the lattice. We therefore see that the Thouless pump in 1D maps onto a quantum Hall problem in 2D!  Indeed the topological invariant in Ref. \cite{TPump} is nothing but the Chern number in the mixed coordinates of spatial momentum and `time-momentum' (phase).
Due to the presence of crossed ``electric" and ``magnetic" fields, a particle placed in the lattice will experience drift, with velocity orthogonal to both, of the magnitude given by their ratio, $v_{drift} = \omega_0/k$. This is nothing but the speed of the potential in the original problem, which yields the pumping result of Thouless.


\bibliography{floq}

\begin{thebibliography}{49}%
\makeatletter
\providecommand \@ifxundefined [1]{%
 \@ifx{#1\undefined}
}%
\providecommand \@ifnum [1]{%
 \ifnum #1\expandafter \@firstoftwo
 \else \expandafter \@secondoftwo
 \fi
}%
\providecommand \@ifx [1]{%
 \ifx #1\expandafter \@firstoftwo
 \else \expandafter \@secondoftwo
 \fi
}%
\providecommand \natexlab [1]{#1}%
\providecommand \enquote  [1]{``#1''}%
\providecommand \bibnamefont  [1]{#1}%
\providecommand \bibfnamefont [1]{#1}%
\providecommand \citenamefont [1]{#1}%
\providecommand \href@noop [0]{\@secondoftwo}%
\providecommand \href [0]{\begingroup \@sanitize@url \@href}%
\providecommand \@href[1]{\@@startlink{#1}\@@href}%
\providecommand \@@href[1]{\endgroup#1\@@endlink}%
\providecommand \@sanitize@url [0]{\catcode `\\12\catcode `\$12\catcode
  `\&12\catcode `\#12\catcode `\^12\catcode `\_12\catcode `\%12\relax}%
\providecommand \@@startlink[1]{}%
\providecommand \@@endlink[0]{}%
\providecommand \url  [0]{\begingroup\@sanitize@url \@url }%
\providecommand \@url [1]{\endgroup\@href {#1}{\urlprefix }}%
\providecommand \urlprefix  [0]{URL }%
\providecommand \Eprint [0]{\href }%
\providecommand \doibase [0]{http://dx.doi.org/}%
\providecommand \selectlanguage [0]{\@gobble}%
\providecommand \bibinfo  [0]{\@secondoftwo}%
\providecommand \bibfield  [0]{\@secondoftwo}%
\providecommand \translation [1]{[#1]}%
\providecommand \BibitemOpen [0]{}%
\providecommand \bibitemStop [0]{}%
\providecommand \bibitemNoStop [0]{.\EOS\space}%
\providecommand \EOS [0]{\spacefactor3000\relax}%
\providecommand \BibitemShut  [1]{\csname bibitem#1\endcsname}%
\let\auto@bib@innerbib\@empty
\bibitem [{\citenamefont {Oka}\ and\ \citenamefont {Aoki}(2009)}]{Oka-09}%
  \BibitemOpen
  \bibfield  {author} {\bibinfo {author} {\bibfnamefont {T.}~\bibnamefont
  {Oka}}\ and\ \bibinfo {author} {\bibfnamefont {H.}~\bibnamefont {Aoki}},\
  }\href {\doibase 10.1103/PhysRevB.79.081406} {\bibfield  {journal} {\bibinfo
  {journal} {Phys. Rev. B}\ }\textbf {\bibinfo {volume} {79}},\ \bibinfo
  {pages} {081406} (\bibinfo {year} {2009})}\BibitemShut {NoStop}%
\bibitem [{\citenamefont {Inoue}\ and\ \citenamefont {Tanaka}(2010)}]{Inoue10}%
  \BibitemOpen
  \bibfield  {author} {\bibinfo {author} {\bibfnamefont {J.-i.}\ \bibnamefont
  {Inoue}}\ and\ \bibinfo {author} {\bibfnamefont {A.}~\bibnamefont {Tanaka}},\
  }\href {\doibase 10.1103/PhysRevLett.105.017401} {\bibfield  {journal}
  {\bibinfo  {journal} {Phys. Rev. Lett.}\ }\textbf {\bibinfo {volume} {105}},\
  \bibinfo {pages} {017401} (\bibinfo {year} {2010})}\BibitemShut {NoStop}%
\bibitem [{\citenamefont {Lindner}\ \emph
  {et~al.}(2011{\natexlab{a}})\citenamefont {Lindner}, \citenamefont {Refael},\
  and\ \citenamefont {Galitski}}]{FTI}%
  \BibitemOpen
  \bibfield  {author} {\bibinfo {author} {\bibfnamefont {N.~H.}\ \bibnamefont
  {Lindner}}, \bibinfo {author} {\bibfnamefont {G.}~\bibnamefont {Refael}}, \
  and\ \bibinfo {author} {\bibfnamefont {V.}~\bibnamefont {Galitski}},\ }\href
  {\doibase 10.1038/nphys1926} {\bibfield  {journal} {\bibinfo  {journal} {Nat
  Phys}\ }\textbf {\bibinfo {volume} {7}},\ \bibinfo {pages} {490} (\bibinfo
  {year} {2011}{\natexlab{a}})}\BibitemShut {NoStop}%
\bibitem [{\citenamefont {Lindner}\ \emph {et~al.}(2013)\citenamefont
  {Lindner}, \citenamefont {Bergman}, \citenamefont {Refael},\ and\
  \citenamefont {Galitski}}]{FTI-3d}%
  \BibitemOpen
  \bibfield  {author} {\bibinfo {author} {\bibfnamefont {N.~H.}\ \bibnamefont
  {Lindner}}, \bibinfo {author} {\bibfnamefont {D.~L.}\ \bibnamefont
  {Bergman}}, \bibinfo {author} {\bibfnamefont {G.}~\bibnamefont {Refael}}, \
  and\ \bibinfo {author} {\bibfnamefont {V.}~\bibnamefont {Galitski}},\ }\href
  {\doibase 10.1103/PhysRevB.87.235131} {\bibfield  {journal} {\bibinfo
  {journal} {Phys. Rev. B}\ }\textbf {\bibinfo {volume} {87}},\ \bibinfo
  {pages} {235131} (\bibinfo {year} {2013})}\BibitemShut {NoStop}%
\bibitem [{\citenamefont {Kitagawa}\ \emph {et~al.}(2011)\citenamefont
  {Kitagawa}, \citenamefont {Oka}, \citenamefont {Brataas}, \citenamefont
  {Fu},\ and\ \citenamefont {Demler}}]{Kitagawa}%
  \BibitemOpen
  \bibfield  {author} {\bibinfo {author} {\bibfnamefont {T.}~\bibnamefont
  {Kitagawa}}, \bibinfo {author} {\bibfnamefont {T.}~\bibnamefont {Oka}},
  \bibinfo {author} {\bibfnamefont {A.}~\bibnamefont {Brataas}}, \bibinfo
  {author} {\bibfnamefont {L.}~\bibnamefont {Fu}}, \ and\ \bibinfo {author}
  {\bibfnamefont {E.}~\bibnamefont {Demler}},\ }\href {\doibase
  10.1103/PhysRevB.84.235108} {\bibfield  {journal} {\bibinfo  {journal} {Phys.
  Rev. B}\ }\textbf {\bibinfo {volume} {84}},\ \bibinfo {pages} {235108}
  (\bibinfo {year} {2011})}\BibitemShut {NoStop}%
\bibitem [{\citenamefont {Rudner}\ \emph {et~al.}(2013)\citenamefont {Rudner},
  \citenamefont {Lindner}, \citenamefont {Berg},\ and\ \citenamefont
  {Levin}}]{AFAI-1}%
  \BibitemOpen
  \bibfield  {author} {\bibinfo {author} {\bibfnamefont {M.~S.}\ \bibnamefont
  {Rudner}}, \bibinfo {author} {\bibfnamefont {N.~H.}\ \bibnamefont {Lindner}},
  \bibinfo {author} {\bibfnamefont {E.}~\bibnamefont {Berg}}, \ and\ \bibinfo
  {author} {\bibfnamefont {M.}~\bibnamefont {Levin}},\ }\href {\doibase
  10.1103/PhysRevX.3.031005} {\bibfield  {journal} {\bibinfo  {journal} {Phys.
  Rev. X}\ }\textbf {\bibinfo {volume} {3}},\ \bibinfo {pages} {031005}
  (\bibinfo {year} {2013})}\BibitemShut {NoStop}%
\bibitem [{\citenamefont {{Titum}}\ \emph {et~al.}(2016)\citenamefont
  {{Titum}}, \citenamefont {{Berg}}, \citenamefont {{Rudner}}, \citenamefont
  {{Refael}},\ and\ \citenamefont {{Lindner}}}]{AFAI-2}%
  \BibitemOpen
  \bibfield  {author} {\bibinfo {author} {\bibfnamefont {P.}~\bibnamefont
  {{Titum}}}, \bibinfo {author} {\bibfnamefont {E.}~\bibnamefont {{Berg}}},
  \bibinfo {author} {\bibfnamefont {M.~S.}\ \bibnamefont {{Rudner}}}, \bibinfo
  {author} {\bibfnamefont {G.}~\bibnamefont {{Refael}}}, \ and\ \bibinfo
  {author} {\bibfnamefont {N.~H.}\ \bibnamefont {{Lindner}}},\ }\href {\doibase
  10.1103/PhysRevX.6.021013} {\bibfield  {journal} {\bibinfo  {journal}
  {Physical Review X}\ }\textbf {\bibinfo {volume} {6}},\ \bibinfo {eid}
  {021013} (\bibinfo {year} {2016})},\ \Eprint
  {http://arxiv.org/abs/1506.00650} {arXiv:1506.00650 [cond-mat.mes-hall]}
  \BibitemShut {NoStop}%
\bibitem [{\citenamefont {Lindner}\ \emph
  {et~al.}(2011{\natexlab{b}})\citenamefont {Lindner}, \citenamefont {Refael},\
  and\ \citenamefont {Galitski}}]{Lindner11}%
  \BibitemOpen
  \bibfield  {author} {\bibinfo {author} {\bibfnamefont {N.~H.}\ \bibnamefont
  {Lindner}}, \bibinfo {author} {\bibfnamefont {G.}~\bibnamefont {Refael}}, \
  and\ \bibinfo {author} {\bibfnamefont {V.}~\bibnamefont {Galitski}},\
  }\href@noop {} {\bibfield  {journal} {\bibinfo  {journal} {Nature Physics}\
  }\textbf {\bibinfo {volume} {7}},\ \bibinfo {pages} {490} (\bibinfo {year}
  {2011}{\natexlab{b}})}\BibitemShut {NoStop}%
\bibitem [{\citenamefont {{Nathan}}\ and\ \citenamefont
  {{Rudner}}(2015)}]{NathanRudner}%
  \BibitemOpen
  \bibfield  {author} {\bibinfo {author} {\bibfnamefont {F.}~\bibnamefont
  {{Nathan}}}\ and\ \bibinfo {author} {\bibfnamefont {M.~S.}\ \bibnamefont
  {{Rudner}}},\ }\href {\doibase 10.1088/1367-2630/17/12/125014} {\bibfield
  {journal} {\bibinfo  {journal} {New Journal of Physics}\ }\textbf {\bibinfo
  {volume} {17}},\ \bibinfo {eid} {125014} (\bibinfo {year} {2015})},\ \Eprint
  {http://arxiv.org/abs/1506.07647} {arXiv:1506.07647 [cond-mat.mes-hall]}
  \BibitemShut {NoStop}%
\bibitem [{\citenamefont {{Roy}}\ and\ \citenamefont
  {{Harper}}(2016{\natexlab{a}})}]{Roy1}%
  \BibitemOpen
  \bibfield  {author} {\bibinfo {author} {\bibfnamefont {R.}~\bibnamefont
  {{Roy}}}\ and\ \bibinfo {author} {\bibfnamefont {F.}~\bibnamefont
  {{Harper}}},\ }\href@noop {} {\bibfield  {journal} {\bibinfo  {journal}
  {ArXiv e-prints}\ } (\bibinfo {year} {2016}{\natexlab{a}})},\ \Eprint
  {http://arxiv.org/abs/1603.06944} {arXiv:1603.06944 [cond-mat.str-el]}
  \BibitemShut {NoStop}%
\bibitem [{\citenamefont {{Roy}}\ and\ \citenamefont
  {{Harper}}(2016{\natexlab{b}})}]{Roy2}%
  \BibitemOpen
  \bibfield  {author} {\bibinfo {author} {\bibfnamefont {R.}~\bibnamefont
  {{Roy}}}\ and\ \bibinfo {author} {\bibfnamefont {F.}~\bibnamefont
  {{Harper}}},\ }\href@noop {} {\bibfield  {journal} {\bibinfo  {journal}
  {ArXiv e-prints}\ } (\bibinfo {year} {2016}{\natexlab{b}})},\ \Eprint
  {http://arxiv.org/abs/1610.06899} {arXiv:1610.06899 [cond-mat.str-el]}
  \BibitemShut {NoStop}%
\bibitem [{\citenamefont {{von Keyserlingk}}\ and\ \citenamefont
  {{Sondhi}}(2016{\natexlab{a}})}]{Sondhi}%
  \BibitemOpen
  \bibfield  {author} {\bibinfo {author} {\bibfnamefont {C.~W.}\ \bibnamefont
  {{von Keyserlingk}}}\ and\ \bibinfo {author} {\bibfnamefont {S.~L.}\
  \bibnamefont {{Sondhi}}},\ }\href {\doibase 10.1103/PhysRevB.93.245145}
  {\bibfield  {journal} {\bibinfo  {journal} {\prb}\ }\textbf {\bibinfo
  {volume} {93}},\ \bibinfo {eid} {245145} (\bibinfo {year}
  {2016}{\natexlab{a}})},\ \Eprint {http://arxiv.org/abs/1602.02157}
  {arXiv:1602.02157 [cond-mat.str-el]} \BibitemShut {NoStop}%
\bibitem [{\citenamefont {{Potter}}\ \emph {et~al.}(2016)\citenamefont
  {{Potter}}, \citenamefont {{Morimoto}},\ and\ \citenamefont
  {{Vishwanath}}}]{potter}%
  \BibitemOpen
  \bibfield  {author} {\bibinfo {author} {\bibfnamefont {A.~C.}\ \bibnamefont
  {{Potter}}}, \bibinfo {author} {\bibfnamefont {T.}~\bibnamefont
  {{Morimoto}}}, \ and\ \bibinfo {author} {\bibfnamefont {A.}~\bibnamefont
  {{Vishwanath}}},\ }\href {\doibase 10.1103/PhysRevX.6.041001} {\bibfield
  {journal} {\bibinfo  {journal} {Physical Review X}\ }\textbf {\bibinfo
  {volume} {6}},\ \bibinfo {eid} {041001} (\bibinfo {year} {2016})}\BibitemShut
  {NoStop}%
\bibitem [{\citenamefont {{von Keyserlingk}}\ and\ \citenamefont
  {{Sondhi}}(2016{\natexlab{b}})}]{TC1}%
  \BibitemOpen
  \bibfield  {author} {\bibinfo {author} {\bibfnamefont {C.~W.}\ \bibnamefont
  {{von Keyserlingk}}}\ and\ \bibinfo {author} {\bibfnamefont {S.~L.}\
  \bibnamefont {{Sondhi}}},\ }\href@noop {} {\bibfield  {journal} {\bibinfo
  {journal} {ArXiv e-prints}\ } (\bibinfo {year} {2016}{\natexlab{b}})},\
  \Eprint {http://arxiv.org/abs/1602.06949} {arXiv:1602.06949
  [cond-mat.str-el]} \BibitemShut {NoStop}%
\bibitem [{\citenamefont {{von Keyserlingk}}\ \emph {et~al.}(2016)\citenamefont
  {{von Keyserlingk}}, \citenamefont {{Khemani}},\ and\ \citenamefont
  {{Sondhi}}}]{TC2}%
  \BibitemOpen
  \bibfield  {author} {\bibinfo {author} {\bibfnamefont {C.~W.}\ \bibnamefont
  {{von Keyserlingk}}}, \bibinfo {author} {\bibfnamefont {V.}~\bibnamefont
  {{Khemani}}}, \ and\ \bibinfo {author} {\bibfnamefont {S.~L.}\ \bibnamefont
  {{Sondhi}}},\ }\href {\doibase 10.1103/PhysRevB.94.085112} {\bibfield
  {journal} {\bibinfo  {journal} {\prb}\ }\textbf {\bibinfo {volume} {94}},\
  \bibinfo {eid} {085112} (\bibinfo {year} {2016})},\ \Eprint
  {http://arxiv.org/abs/1605.00639} {arXiv:1605.00639 [cond-mat.dis-nn]}
  \BibitemShut {NoStop}%
\bibitem [{\citenamefont {{Else}}\ \emph {et~al.}(2016)\citenamefont {{Else}},
  \citenamefont {{Bauer}},\ and\ \citenamefont {{Nayak}}}]{TC4}%
  \BibitemOpen
  \bibfield  {author} {\bibinfo {author} {\bibfnamefont {D.~V.}\ \bibnamefont
  {{Else}}}, \bibinfo {author} {\bibfnamefont {B.}~\bibnamefont {{Bauer}}}, \
  and\ \bibinfo {author} {\bibfnamefont {C.}~\bibnamefont {{Nayak}}},\
  }\href@noop {} {\bibfield  {journal} {\bibinfo  {journal} {ArXiv e-prints}\ }
  (\bibinfo {year} {2016})},\ \Eprint {http://arxiv.org/abs/1607.05277}
  {arXiv:1607.05277 [cond-mat.stat-mech]} \BibitemShut {NoStop}%
\bibitem [{\citenamefont {{Potirniche}}\ \emph {et~al.}(2016)\citenamefont
  {{Potirniche}}, \citenamefont {{Potter}}, \citenamefont {{Schleier-Smith}},
  \citenamefont {{Vishwanath}},\ and\ \citenamefont {{Yao}}}]{TC5}%
  \BibitemOpen
  \bibfield  {author} {\bibinfo {author} {\bibfnamefont {I.-D.}\ \bibnamefont
  {{Potirniche}}}, \bibinfo {author} {\bibfnamefont {A.~C.}\ \bibnamefont
  {{Potter}}}, \bibinfo {author} {\bibfnamefont {M.}~\bibnamefont
  {{Schleier-Smith}}}, \bibinfo {author} {\bibfnamefont {A.}~\bibnamefont
  {{Vishwanath}}}, \ and\ \bibinfo {author} {\bibfnamefont {N.~Y.}\
  \bibnamefont {{Yao}}},\ }\href@noop {} {\bibfield  {journal} {\bibinfo
  {journal} {ArXiv e-prints}\ } (\bibinfo {year} {2016})},\ \Eprint
  {http://arxiv.org/abs/1610.07611} {arXiv:1610.07611 [cond-mat.quant-gas]}
  \BibitemShut {NoStop}%
\bibitem [{\citenamefont {{Abanin}}\ \emph {et~al.}(2014)\citenamefont
  {{Abanin}}, \citenamefont {{De Roeck}},\ and\ \citenamefont
  {{Huveneers}}}]{Abanin}%
  \BibitemOpen
  \bibfield  {author} {\bibinfo {author} {\bibfnamefont {D.}~\bibnamefont
  {{Abanin}}}, \bibinfo {author} {\bibfnamefont {W.}~\bibnamefont {{De
  Roeck}}}, \ and\ \bibinfo {author} {\bibfnamefont {F.}~\bibnamefont
  {{Huveneers}}},\ }\href@noop {} {\bibfield  {journal} {\bibinfo  {journal}
  {ArXiv e-prints}\ } (\bibinfo {year} {2014})},\ \Eprint
  {http://arxiv.org/abs/1412.4752} {arXiv:1412.4752 [cond-mat.dis-nn]}
  \BibitemShut {NoStop}%
\bibitem [{\citenamefont {{Lazarides}}\ \emph {et~al.}(2015)\citenamefont
  {{Lazarides}}, \citenamefont {{Das}},\ and\ \citenamefont
  {{Moessner}}}]{Roderich-FL}%
  \BibitemOpen
  \bibfield  {author} {\bibinfo {author} {\bibfnamefont {A.}~\bibnamefont
  {{Lazarides}}}, \bibinfo {author} {\bibfnamefont {A.}~\bibnamefont {{Das}}},
  \ and\ \bibinfo {author} {\bibfnamefont {R.}~\bibnamefont {{Moessner}}},\
  }\href {\doibase 10.1103/PhysRevLett.115.030402} {\bibfield  {journal}
  {\bibinfo  {journal} {Physical Review Letters}\ }\textbf {\bibinfo {volume}
  {115}},\ \bibinfo {eid} {030402} (\bibinfo {year} {2015})},\ \Eprint
  {http://arxiv.org/abs/1410.3455} {arXiv:1410.3455 [cond-mat.stat-mech]}
  \BibitemShut {NoStop}%
\bibitem [{\citenamefont {{Zhang}}\ \emph
  {et~al.}(2016{\natexlab{a}})\citenamefont {{Zhang}}, \citenamefont
  {{Khemani}},\ and\ \citenamefont {{Huse}}}]{Huse}%
  \BibitemOpen
  \bibfield  {author} {\bibinfo {author} {\bibfnamefont {L.}~\bibnamefont
  {{Zhang}}}, \bibinfo {author} {\bibfnamefont {V.}~\bibnamefont {{Khemani}}},
  \ and\ \bibinfo {author} {\bibfnamefont {D.~A.}\ \bibnamefont {{Huse}}},\
  }\href@noop {} {\bibfield  {journal} {\bibinfo  {journal} {ArXiv e-prints}\ }
  (\bibinfo {year} {2016}{\natexlab{a}})},\ \Eprint
  {http://arxiv.org/abs/1609.00390} {arXiv:1609.00390 [cond-mat.dis-nn]}
  \BibitemShut {NoStop}%
\bibitem [{\citenamefont {Wang}\ \emph {et~al.}(2013)\citenamefont {Wang},
  \citenamefont {Steinberg}, \citenamefont {Jarillo-Herrero},\ and\
  \citenamefont {Gedik}}]{Gedik}%
  \BibitemOpen
  \bibfield  {author} {\bibinfo {author} {\bibfnamefont {Y.~H.}\ \bibnamefont
  {Wang}}, \bibinfo {author} {\bibfnamefont {H.}~\bibnamefont {Steinberg}},
  \bibinfo {author} {\bibfnamefont {P.}~\bibnamefont {Jarillo-Herrero}}, \ and\
  \bibinfo {author} {\bibfnamefont {N.}~\bibnamefont {Gedik}},\ }\href
  {\doibase 10.1126/science.1239834} {\bibfield  {journal} {\bibinfo  {journal}
  {Science}\ }\textbf {\bibinfo {volume} {342}},\ \bibinfo {pages} {453}
  (\bibinfo {year} {2013})}\BibitemShut {NoStop}%
\bibitem [{\citenamefont {Rechtsman}\ \emph {et~al.}(2013)\citenamefont
  {Rechtsman}, \citenamefont {Zeuner}, \citenamefont {Plotnik}, \citenamefont
  {Lumer}, \citenamefont {Podolsky}, \citenamefont {Dreisow}, \citenamefont
  {Nolte}, \citenamefont {Segev},\ and\ \citenamefont
  {Szameit}}]{Rechtsman2013}%
  \BibitemOpen
  \bibfield  {author} {\bibinfo {author} {\bibfnamefont {M.~C.}\ \bibnamefont
  {Rechtsman}}, \bibinfo {author} {\bibfnamefont {J.~M.}\ \bibnamefont
  {Zeuner}}, \bibinfo {author} {\bibfnamefont {Y.}~\bibnamefont {Plotnik}},
  \bibinfo {author} {\bibfnamefont {Y.}~\bibnamefont {Lumer}}, \bibinfo
  {author} {\bibfnamefont {D.}~\bibnamefont {Podolsky}}, \bibinfo {author}
  {\bibfnamefont {F.}~\bibnamefont {Dreisow}}, \bibinfo {author} {\bibfnamefont
  {S.}~\bibnamefont {Nolte}}, \bibinfo {author} {\bibfnamefont
  {M.}~\bibnamefont {Segev}}, \ and\ \bibinfo {author} {\bibfnamefont
  {A.}~\bibnamefont {Szameit}},\ }\href@noop {} {\bibfield  {journal} {\bibinfo
   {journal} {Nature}\ }\textbf {\bibinfo {volume} {496}},\ \bibinfo {pages}
  {196} (\bibinfo {year} {2013})},\ \bibinfo {note} {letter}\BibitemShut
  {NoStop}%
\bibitem [{\citenamefont {{Zhang}}\ \emph
  {et~al.}(2016{\natexlab{b}})\citenamefont {{Zhang}}, \citenamefont {{Hess}},
  \citenamefont {{Kyprianidis}}, \citenamefont {{Becker}}, \citenamefont
  {{Lee}}, \citenamefont {{Smith}}, \citenamefont {{Pagano}}, \citenamefont
  {{Potirniche}}, \citenamefont {{Potter}}, \citenamefont {{Vishwanath}},
  \citenamefont {{Yao}},\ and\ \citenamefont {{Monroe}}}]{Monroe}%
  \BibitemOpen
  \bibfield  {author} {\bibinfo {author} {\bibfnamefont {J.}~\bibnamefont
  {{Zhang}}}, \bibinfo {author} {\bibfnamefont {P.~W.}\ \bibnamefont {{Hess}}},
  \bibinfo {author} {\bibfnamefont {A.}~\bibnamefont {{Kyprianidis}}}, \bibinfo
  {author} {\bibfnamefont {P.}~\bibnamefont {{Becker}}}, \bibinfo {author}
  {\bibfnamefont {A.}~\bibnamefont {{Lee}}}, \bibinfo {author} {\bibfnamefont
  {J.}~\bibnamefont {{Smith}}}, \bibinfo {author} {\bibfnamefont
  {G.}~\bibnamefont {{Pagano}}}, \bibinfo {author} {\bibfnamefont {I.-D.}\
  \bibnamefont {{Potirniche}}}, \bibinfo {author} {\bibfnamefont {A.~C.}\
  \bibnamefont {{Potter}}}, \bibinfo {author} {\bibfnamefont {A.}~\bibnamefont
  {{Vishwanath}}}, \bibinfo {author} {\bibfnamefont {N.~Y.}\ \bibnamefont
  {{Yao}}}, \ and\ \bibinfo {author} {\bibfnamefont {C.}~\bibnamefont
  {{Monroe}}},\ }\href@noop {} {\bibfield  {journal} {\bibinfo  {journal}
  {ArXiv e-prints}\ } (\bibinfo {year} {2016}{\natexlab{b}})},\ \Eprint
  {http://arxiv.org/abs/1609.08684} {arXiv:1609.08684 [quant-ph]} \BibitemShut
  {NoStop}%
\bibitem [{\citenamefont {{Kalugin}}\ \emph {et~al.}(1985)\citenamefont
  {{Kalugin}}, \citenamefont {{Kitaev}},\ and\ \citenamefont
  {{Levitov}}}]{KitaevLevitov}%
  \BibitemOpen
  \bibfield  {author} {\bibinfo {author} {\bibfnamefont {P.~A.}\ \bibnamefont
  {{Kalugin}}}, \bibinfo {author} {\bibfnamefont {A.~I.}\ \bibnamefont
  {{Kitaev}}}, \ and\ \bibinfo {author} {\bibfnamefont {L.~S.}\ \bibnamefont
  {{Levitov}}},\ }\href@noop {} {\bibfield  {journal} {\bibinfo  {journal}
  {ZhETF Pisma Redaktsiiu}\ }\textbf {\bibinfo {volume} {41}},\ \bibinfo
  {pages} {119} (\bibinfo {year} {1985})}\BibitemShut {NoStop}%
\bibitem [{\citenamefont {{Levine}}\ and\ \citenamefont
  {{Steinhardt}}(1984)}]{LevineSteinhardt}%
  \BibitemOpen
  \bibfield  {author} {\bibinfo {author} {\bibfnamefont {D.}~\bibnamefont
  {{Levine}}}\ and\ \bibinfo {author} {\bibfnamefont {P.~J.}\ \bibnamefont
  {{Steinhardt}}},\ }\href {\doibase 10.1103/PhysRevLett.53.2477} {\bibfield
  {journal} {\bibinfo  {journal} {Physical Review Letters}\ }\textbf {\bibinfo
  {volume} {53}},\ \bibinfo {pages} {2477} (\bibinfo {year}
  {1984})}\BibitemShut {NoStop}%
\bibitem [{\citenamefont {Shechtman}\ \emph {et~al.}(1984)\citenamefont
  {Shechtman}, \citenamefont {Blech}, \citenamefont {Gratias},\ and\
  \citenamefont {Cahn}}]{Shechtman}%
  \BibitemOpen
  \bibfield  {author} {\bibinfo {author} {\bibfnamefont {D.}~\bibnamefont
  {Shechtman}}, \bibinfo {author} {\bibfnamefont {I.}~\bibnamefont {Blech}},
  \bibinfo {author} {\bibfnamefont {D.}~\bibnamefont {Gratias}}, \ and\
  \bibinfo {author} {\bibfnamefont {J.~W.}\ \bibnamefont {Cahn}},\ }\href
  {\doibase 10.1103/PhysRevLett.53.1951} {\bibfield  {journal} {\bibinfo
  {journal} {Phys. Rev. Lett.}\ }\textbf {\bibinfo {volume} {53}},\ \bibinfo
  {pages} {1951} (\bibinfo {year} {1984})}\BibitemShut {NoStop}%
\bibitem [{\citenamefont {Bernevig}\ \emph {et~al.}(2006)\citenamefont
  {Bernevig}, \citenamefont {Hughes},\ and\ \citenamefont
  {Zhang}}]{Bernevig06}%
  \BibitemOpen
  \bibfield  {author} {\bibinfo {author} {\bibfnamefont {B.~A.}\ \bibnamefont
  {Bernevig}}, \bibinfo {author} {\bibfnamefont {T.~L.}\ \bibnamefont
  {Hughes}}, \ and\ \bibinfo {author} {\bibfnamefont {S.-C.}\ \bibnamefont
  {Zhang}},\ }\href {\doibase 10.1126/science.1133734} {\bibfield  {journal}
  {\bibinfo  {journal} {Science}\ }\textbf {\bibinfo {volume} {314}},\ \bibinfo
  {pages} {1757} (\bibinfo {year} {2006})},\ \Eprint
  {http://arxiv.org/abs/http://science.sciencemag.org/content/314/5806/1757.full.pdf}
  {http://science.sciencemag.org/content/314/5806/1757.full.pdf} \BibitemShut
  {NoStop}%
\bibitem [{\citenamefont {Mukamel}(1999)}]{Muka}%
  \BibitemOpen
  \bibfield  {author} {\bibinfo {author} {\bibfnamefont {S.}~\bibnamefont
  {Mukamel}},\ }\href@noop {} {\emph {\bibinfo {title} {Principles of nonlinear
  optical spectroscopy}}},\ \bibinfo {number} {6}\ (\bibinfo  {publisher}
  {Oxford University Press on Demand},\ \bibinfo {year} {1999})\BibitemShut
  {NoStop}%
\bibitem [{\citenamefont {Floquet}(1883)}]{Floquet}%
  \BibitemOpen
  \bibfield  {author} {\bibinfo {author} {\bibfnamefont {G.}~\bibnamefont
  {Floquet}},\ }in\ \href@noop {} {\emph {\bibinfo {booktitle} {Annales
  scientifiques de l'{\'E}cole normale sup{\'e}rieure}}},\ Vol.~\bibinfo
  {volume} {12}\ (\bibinfo {year} {1883})\ pp.\ \bibinfo {pages}
  {47--88}\BibitemShut {NoStop}%
\bibitem [{\citenamefont {Bloch}(1929)}]{Bloch}%
  \BibitemOpen
  \bibfield  {author} {\bibinfo {author} {\bibfnamefont {F.}~\bibnamefont
  {Bloch}},\ }\href@noop {} {\bibfield  {journal} {\bibinfo  {journal}
  {Zeitschrift f{\"u}r physik}\ }\textbf {\bibinfo {volume} {52}},\ \bibinfo
  {pages} {555} (\bibinfo {year} {1929})}\BibitemShut {NoStop}%
\bibitem [{\citenamefont {Ho}\ \emph {et~al.}(1983)\citenamefont {Ho},
  \citenamefont {Chu},\ and\ \citenamefont {Tietz}}]{Ho83}%
  \BibitemOpen
  \bibfield  {author} {\bibinfo {author} {\bibfnamefont {T.-S.}\ \bibnamefont
  {Ho}}, \bibinfo {author} {\bibfnamefont {S.-I.}\ \bibnamefont {Chu}}, \ and\
  \bibinfo {author} {\bibfnamefont {J.~V.}\ \bibnamefont {Tietz}},\ }\href
  {\doibase http://dx.doi.org/10.1016/0009-2614(83)80732-5} {\bibfield
  {journal} {\bibinfo  {journal} {Chemical Physics Letters}\ }\textbf {\bibinfo
  {volume} {96}},\ \bibinfo {pages} {464 } (\bibinfo {year}
  {1983})}\BibitemShut {NoStop}%
\bibitem [{\citenamefont {Casati}\ \emph {et~al.}(1989)\citenamefont {Casati},
  \citenamefont {Guarneri},\ and\ \citenamefont {Shepelyansky}}]{Shep89}%
  \BibitemOpen
  \bibfield  {author} {\bibinfo {author} {\bibfnamefont {G.}~\bibnamefont
  {Casati}}, \bibinfo {author} {\bibfnamefont {I.}~\bibnamefont {Guarneri}}, \
  and\ \bibinfo {author} {\bibfnamefont {D.~L.}\ \bibnamefont {Shepelyansky}},\
  }\href {\doibase 10.1103/PhysRevLett.62.345} {\bibfield  {journal} {\bibinfo
  {journal} {Phys. Rev. Lett.}\ }\textbf {\bibinfo {volume} {62}},\ \bibinfo
  {pages} {345} (\bibinfo {year} {1989})}\BibitemShut {NoStop}%
\bibitem [{Note1()}]{Note1}%
  \BibitemOpen
  \bibinfo {note} {It is important here that $\omega _1/\omega _2$ is
  irrational; for rational case see below}\BibitemShut {NoStop}%
\bibitem [{\citenamefont {Nakanishi}\ \emph {et~al.}(1993)\citenamefont
  {Nakanishi}, \citenamefont {Ohtsuki},\ and\ \citenamefont
  {Saitoh}}]{Nakanishi93}%
  \BibitemOpen
  \bibfield  {author} {\bibinfo {author} {\bibfnamefont {T.}~\bibnamefont
  {Nakanishi}}, \bibinfo {author} {\bibfnamefont {T.}~\bibnamefont {Ohtsuki}},
  \ and\ \bibinfo {author} {\bibfnamefont {M.}~\bibnamefont {Saitoh}},\ }\href
  {\doibase 10.1143/JPSJ.62.2773} {\bibfield  {journal} {\bibinfo  {journal}
  {Journal of the Physical Society of Japan}\ }\textbf {\bibinfo {volume}
  {62}},\ \bibinfo {pages} {2773} (\bibinfo {year} {1993})},\ \Eprint
  {http://arxiv.org/abs/http://dx.doi.org/10.1143/JPSJ.62.2773}
  {http://dx.doi.org/10.1143/JPSJ.62.2773} \BibitemShut {NoStop}%
\bibitem [{\citenamefont {Sundaram}\ and\ \citenamefont {Niu}(1999)}]{Niu99}%
  \BibitemOpen
  \bibfield  {author} {\bibinfo {author} {\bibfnamefont {G.}~\bibnamefont
  {Sundaram}}\ and\ \bibinfo {author} {\bibfnamefont {Q.}~\bibnamefont {Niu}},\
  }\href {\doibase 10.1103/PhysRevB.59.14915} {\bibfield  {journal} {\bibinfo
  {journal} {Phys. Rev. B}\ }\textbf {\bibinfo {volume} {59}},\ \bibinfo
  {pages} {14915} (\bibinfo {year} {1999})}\BibitemShut {NoStop}%
\bibitem [{\citenamefont {Abragam}(1961)}]{abragam1961principles}%
  \BibitemOpen
  \bibfield  {author} {\bibinfo {author} {\bibfnamefont {A.}~\bibnamefont
  {Abragam}},\ }\href@noop {} {\emph {\bibinfo {title} {The principles of
  nuclear magnetism}}},\ \bibinfo {number} {32}\ (\bibinfo  {publisher} {Oxford
  university press},\ \bibinfo {year} {1961})\BibitemShut {NoStop}%
\bibitem [{\citenamefont {Spencer}\ \emph {et~al.}(1959)\citenamefont
  {Spencer}, \citenamefont {LeCraw},\ and\ \citenamefont {Clogston}}]{YIG}%
  \BibitemOpen
  \bibfield  {author} {\bibinfo {author} {\bibfnamefont {E.~G.}\ \bibnamefont
  {Spencer}}, \bibinfo {author} {\bibfnamefont {R.~C.}\ \bibnamefont {LeCraw}},
  \ and\ \bibinfo {author} {\bibfnamefont {A.~M.}\ \bibnamefont {Clogston}},\
  }\href {\doibase 10.1103/PhysRevLett.3.32} {\bibfield  {journal} {\bibinfo
  {journal} {Phys. Rev. Lett.}\ }\textbf {\bibinfo {volume} {3}},\ \bibinfo
  {pages} {32} (\bibinfo {year} {1959})}\BibitemShut {NoStop}%
\bibitem [{\citenamefont {Tyryshkin}\ \emph {et~al.}(2003)\citenamefont
  {Tyryshkin}, \citenamefont {Lyon}, \citenamefont {Astashkin},\ and\
  \citenamefont {Raitsimring}}]{Si:P}%
  \BibitemOpen
  \bibfield  {author} {\bibinfo {author} {\bibfnamefont {A.~M.}\ \bibnamefont
  {Tyryshkin}}, \bibinfo {author} {\bibfnamefont {S.~A.}\ \bibnamefont {Lyon}},
  \bibinfo {author} {\bibfnamefont {A.~V.}\ \bibnamefont {Astashkin}}, \ and\
  \bibinfo {author} {\bibfnamefont {A.~M.}\ \bibnamefont {Raitsimring}},\
  }\href {\doibase 10.1103/PhysRevB.68.193207} {\bibfield  {journal} {\bibinfo
  {journal} {Phys. Rev. B}\ }\textbf {\bibinfo {volume} {68}},\ \bibinfo
  {pages} {193207} (\bibinfo {year} {2003})}\BibitemShut {NoStop}%
\bibitem [{\citenamefont {Manucharyan}\ \emph {et~al.}(2009)\citenamefont
  {Manucharyan}, \citenamefont {Koch}, \citenamefont {Glazman},\ and\
  \citenamefont {Devoret}}]{Manucharyan113}%
  \BibitemOpen
  \bibfield  {author} {\bibinfo {author} {\bibfnamefont {V.~E.}\ \bibnamefont
  {Manucharyan}}, \bibinfo {author} {\bibfnamefont {J.}~\bibnamefont {Koch}},
  \bibinfo {author} {\bibfnamefont {L.~I.}\ \bibnamefont {Glazman}}, \ and\
  \bibinfo {author} {\bibfnamefont {M.~H.}\ \bibnamefont {Devoret}},\ }\href
  {\doibase 10.1126/science.1175552} {\bibfield  {journal} {\bibinfo  {journal}
  {Science}\ }\textbf {\bibinfo {volume} {326}},\ \bibinfo {pages} {113}
  (\bibinfo {year} {2009})},\ \Eprint
  {http://arxiv.org/abs/http://science.sciencemag.org/content/326/5949/113.full.pdf}
  {http://science.sciencemag.org/content/326/5949/113.full.pdf} \BibitemShut
  {NoStop}%
\bibitem [{\citenamefont {van Heck}\ \emph {et~al.}(2014)\citenamefont {van
  Heck}, \citenamefont {Mi},\ and\ \citenamefont {Akhmerov}}]{Heck}%
  \BibitemOpen
  \bibfield  {author} {\bibinfo {author} {\bibfnamefont {B.}~\bibnamefont {van
  Heck}}, \bibinfo {author} {\bibfnamefont {S.}~\bibnamefont {Mi}}, \ and\
  \bibinfo {author} {\bibfnamefont {A.~R.}\ \bibnamefont {Akhmerov}},\ }\href
  {\doibase 10.1103/PhysRevB.90.155450} {\bibfield  {journal} {\bibinfo
  {journal} {Phys. Rev. B}\ }\textbf {\bibinfo {volume} {90}},\ \bibinfo
  {pages} {155450} (\bibinfo {year} {2014})}\BibitemShut {NoStop}%
\bibitem [{\citenamefont {Riwar}\ \emph {et~al.}(2016)\citenamefont {Riwar},
  \citenamefont {Houzet}, \citenamefont {Meyer},\ and\ \citenamefont
  {Nazarov}}]{Meyer}%
  \BibitemOpen
  \bibfield  {author} {\bibinfo {author} {\bibfnamefont {R.-P.}\ \bibnamefont
  {Riwar}}, \bibinfo {author} {\bibfnamefont {M.}~\bibnamefont {Houzet}},
  \bibinfo {author} {\bibfnamefont {J.~S.}\ \bibnamefont {Meyer}}, \ and\
  \bibinfo {author} {\bibfnamefont {Y.~V.}\ \bibnamefont {Nazarov}},\
  }\href@noop {} {\bibfield  {journal} {\bibinfo  {journal} {Nature
  communications}\ }\textbf {\bibinfo {volume} {7}} (\bibinfo {year}
  {2016})}\BibitemShut {NoStop}%
\bibitem [{\citenamefont {Fu}\ \emph {et~al.}(2007)\citenamefont {Fu},
  \citenamefont {Kane},\ and\ \citenamefont {Mele}}]{FKM}%
  \BibitemOpen
  \bibfield  {author} {\bibinfo {author} {\bibfnamefont {L.}~\bibnamefont
  {Fu}}, \bibinfo {author} {\bibfnamefont {C.~L.}\ \bibnamefont {Kane}}, \ and\
  \bibinfo {author} {\bibfnamefont {E.~J.}\ \bibnamefont {Mele}},\ }\href
  {\doibase 10.1103/PhysRevLett.98.106803} {\bibfield  {journal} {\bibinfo
  {journal} {Phys. Rev. Lett.}\ }\textbf {\bibinfo {volume} {98}},\ \bibinfo
  {pages} {106803} (\bibinfo {year} {2007})}\BibitemShut {NoStop}%
\bibitem [{\citenamefont {Li}\ and\ \citenamefont {Wu}(2013)}]{Congjun1}%
  \BibitemOpen
  \bibfield  {author} {\bibinfo {author} {\bibfnamefont {Y.}~\bibnamefont
  {Li}}\ and\ \bibinfo {author} {\bibfnamefont {C.}~\bibnamefont {Wu}},\ }\href
  {\doibase 10.1103/PhysRevLett.110.216802} {\bibfield  {journal} {\bibinfo
  {journal} {Phys. Rev. Lett.}\ }\textbf {\bibinfo {volume} {110}},\ \bibinfo
  {pages} {216802} (\bibinfo {year} {2013})}\BibitemShut {NoStop}%
\bibitem [{\citenamefont {Li}\ \emph {et~al.}(2013)\citenamefont {Li},
  \citenamefont {Zhang},\ and\ \citenamefont {Wu}}]{Congjun2}%
  \BibitemOpen
  \bibfield  {author} {\bibinfo {author} {\bibfnamefont {Y.}~\bibnamefont
  {Li}}, \bibinfo {author} {\bibfnamefont {S.-C.}\ \bibnamefont {Zhang}}, \
  and\ \bibinfo {author} {\bibfnamefont {C.}~\bibnamefont {Wu}},\ }\href
  {\doibase 10.1103/PhysRevLett.111.186803} {\bibfield  {journal} {\bibinfo
  {journal} {Phys. Rev. Lett.}\ }\textbf {\bibinfo {volume} {111}},\ \bibinfo
  {pages} {186803} (\bibinfo {year} {2013})}\BibitemShut {NoStop}%
\bibitem [{\citenamefont {Bernevig}\ \emph {et~al.}(2003)\citenamefont
  {Bernevig}, \citenamefont {Hu}, \citenamefont {Toumbas},\ and\ \citenamefont
  {Zhang}}]{8dHall}%
  \BibitemOpen
  \bibfield  {author} {\bibinfo {author} {\bibfnamefont {B.~A.}\ \bibnamefont
  {Bernevig}}, \bibinfo {author} {\bibfnamefont {J.}~\bibnamefont {Hu}},
  \bibinfo {author} {\bibfnamefont {N.}~\bibnamefont {Toumbas}}, \ and\
  \bibinfo {author} {\bibfnamefont {S.-C.}\ \bibnamefont {Zhang}},\ }\href
  {\doibase 10.1103/PhysRevLett.91.236803} {\bibfield  {journal} {\bibinfo
  {journal} {Phys. Rev. Lett.}\ }\textbf {\bibinfo {volume} {91}},\ \bibinfo
  {pages} {236803} (\bibinfo {year} {2003})}\BibitemShut {NoStop}%
\bibitem [{\citenamefont {Janot}(1994)}]{Janot}%
  \BibitemOpen
  \bibfield  {author} {\bibinfo {author} {\bibfnamefont {C.}~\bibnamefont
  {Janot}},\ }\href@noop {} {\emph {\bibinfo {title} {Quasicrystals}}}\
  (\bibinfo  {publisher} {Springer},\ \bibinfo {year} {1994})\BibitemShut
  {NoStop}%
\bibitem [{\citenamefont {{Hurwitz}}(1891)}]{Hurwitz}%
  \BibitemOpen
  \bibfield  {author} {\bibinfo {author} {\bibfnamefont {A.}~\bibnamefont
  {{Hurwitz}}},\ }\href {\doibase 10.1007/BF01206656} {\bibfield  {journal}
  {\bibinfo  {journal} {{Math. Ann.}}\ }\textbf {\bibinfo {volume} {39}},\
  \bibinfo {pages} {279} (\bibinfo {year} {1891})}\BibitemShut {NoStop}%
\bibitem [{\citenamefont {Brouwer}(1998)}]{Brouwer}%
  \BibitemOpen
  \bibfield  {author} {\bibinfo {author} {\bibfnamefont {P.~W.}\ \bibnamefont
  {Brouwer}},\ }\href {\doibase 10.1103/PhysRevB.58.R10135} {\bibfield
  {journal} {\bibinfo  {journal} {Phys. Rev. B}\ }\textbf {\bibinfo {volume}
  {58}},\ \bibinfo {pages} {R10135} (\bibinfo {year} {1998})}\BibitemShut
  {NoStop}%
\bibitem [{\citenamefont {Thouless}(1983)}]{TPump}%
  \BibitemOpen
  \bibfield  {author} {\bibinfo {author} {\bibfnamefont {D.~J.}\ \bibnamefont
  {Thouless}},\ }\href {\doibase 10.1103/PhysRevB.27.6083} {\bibfield
  {journal} {\bibinfo  {journal} {Phys. Rev. B}\ }\textbf {\bibinfo {volume}
  {27}},\ \bibinfo {pages} {6083} (\bibinfo {year} {1983})}\BibitemShut
             {NoStop}%
\bibitem [75]{HBD}%
\BibitemOpen
\bibfield {author}{\bibinfo{author}{\bibnamefont { B.~I.~Halperin,~}}}{\bibfield {howpublished}{\bibinfo {howpublished} {hbd (1941)}}}
      \BibitemShut
          {NoStop}%
\end{thebibliography}%


\end{document}